# Green Haven or Risky Venture? Exploring the Connectedness and Hedging of Sustainable Cryptocurrencies and Green Financial Markets

Chang Li[1], Rui Jiang[2], Hao Wu[3], Conghua Wen[4, *]

## Abstract

Conventional cryptocurrency often leads to increased energy consumption and carbon emissions, while sustainable cryptocurrencies possess the potential to become a green alternative in portfolio management. This study aims to investigate the time-varying connectedness between sustainable cryptocurrency and green financial markets as well as hedging performance when facing market shocks, including COVID-19 and Russia-Ukraine war. TVP-VAR model with Fourier transform and Multivariate GARCH models are employed. The findings indicate that the pairwise connectedness between the sustainable cryptocurrencies and green financial markets has been at a low level, providing diversification benefits in investment portfolio. Besides, short-term connectedness dominates medium- and long-term connectedness. Sustainable cryptocurrencies show higher hedging effectiveness than traditional cryptocurrency.

---

[1] School of Mathematics and Physics, Xi'an Jiaotong-Liverpool University, China. Email: chang.li18@student.xjtlu.edu.cn.

[2] School of Information Management and Engineering, Shanghai University of Finance and Economics, China. Email: rui.jiang@163.sufe.edu.cn.

[3] School of Economics and Management, Nanjing Institute of Technology, China, Email: wuhao@njit.edu.cn.

[4] School of Mathematics and Physics, Xi'an Jiaotong-Liverpool University, China. Email: Conghua.Wen@xjtlu.edu.cn.

* Corresponding author.

# 1. Introduction

In the context of significant fluctuations in traditional financial markets due to geopolitical conflicts and global economic uncertainty, many investors have sought refuge in cryptocurrencies as a safe haven, thereby showing their potential for portfolio diversification. However, their mining also sparks an increasing number of discussions about environmental impact and sustainability (Petukhina et al., 2021; Rao et al., 2022). Cryptocurrency mining requires noticeable computing power, and the calculation occupies a large number of specialized hardware devices, which usually involves the use of fossil fuels, and produces greenhouse gases (Proelss et al., 2023; Li and Meng, 2022). For instance, as of February 2025, the annualized total electricity consumption of Bitcoin was 175.87 terawatt-hours, equivalent to that of Poland, and its annualized total carbon footprint was 98.10 million metric tons of carbon dioxide, comparable to that of Qatar.[5] Given these concerns, some sustainable cryptocurrency projects are considering or have adopted different consensus mechanisms that do not require significant computing power and energy consumption (Sharif et al., 2023). Recent studies have categorized the cryptocurrencies into traditional and sustainable cryptocurrencies based on their potential environmental impacts, instead of regarding them as a homogeneous entity (Pham et al. 2022; Ali et al., 2024).

Meanwhile, geopolitical conflicts prompt a global reassessment of energy independence and security, leading to an increase in investment and support for green energy projects in many countries. Global climate change also raises the awareness of sustainable development and green technologies, and further increased demand for green financial products (Taghizadeh-Hesary and

[5] https://digiconomist.net/bitcoin-energy-consumption

Yoshino, 2019; Lenzen and Dey, 2002; Cumming et al., 2024). Therefore, current studies pay attention to the time-varying trends and financial contagion in green financial markets, offering guidance for portfolio management. Reboredo and Ugolini (2020) found that the green bond market received considerable price spillover from fixed income and money markets. Lu et al. (2023) found S&P Global Clean Energy index is the man main net transmitter of return spillovers in the network consisting of green bonds, clean energy, and socially responsible stocks.

Despite the growing attention to the green financial assets or cryptocurrencies, the intersection of sustainable cryptocurrencies with green financial markets remains largely unexplored. Current studies mainly contribute to the investigation of the interdependence and spillover effect around green financial assets or cryptocurrencies (Ren and Lucey, 2022; Sharif et al., 2023; Siddique et al., 2024). There is a notable gap in the literature regarding the potential of sustainable cryptocurrencies to influence investment strategies and portfolio composition within green financial markets, and the dependence or dynamic hedge ratios between sustainable cryptocurrencies and mainstream green instruments remain largely unexplored. Considering the safe-haven function of cryptocurrencies under the market uncertainty and the profitability of green financial products, it is practical to explore the hedging potential between two kinds of assets. This research aims to address this gap by exploring how sustainable cryptocurrencies can be integrated into green investment portfolios, thereby providing investors with actionable insights to enhance both financial returns and environmental impact. Besides, the emergence of sustainable cryptocurrencies, which prioritize environmental friendliness through innovative consensus mechanisms and renewable energy sources, presents an intriguing opportunity for investors seeking to reconcile their financial goals with environmental responsibilities (Jacquet and Mans, 2019). This study aims to address this research gap by exploring the intersection of sustainable cryptocurrencies and green financial market, with a focus on understanding their combined impact on portfolio sustainability and risk-return profiles.

Therefore, by considering both conventional and sustainable cryptocurrencies, we study the return and volatility connectedness between cryptocurrencies and green financial markets. Moreover, based on the analysis of connectedness, we construct the optimal portfolios and evaluate the diversification opportunities between the two markets. Besides, we observe the variation of information spillover in different time periods, especially under the impact of the COVID-19 and Russia-Ukraine war, which has increased economic uncertainty worldwide. In response, investors have diversified their holdings to mitigate financial risk. Specifically, the Time-Varying Parameter Vector Autoregressive (TVP-VAR) model with Fourier transform is employed to present a comprehensive and novel empirical depiction of the dynamic correlation between the two markets. Given the potential insights into the market interdependencies, this study aims to bridge this gap and offer recommendations to policymakers and investors for assessing this dependency structure and portfolio risk management.

This study contributes to the existing literature in various aspects. It is the first to focus on the portfolio diversification and hedging between sustainable cryptocurrencies and green financial markets under the market shocks including COVID-19 and Russia-Ukraine war. We explore the diverse characteristics between conventional cryptocurrency, and green alternatives, rather than considering them as a homogeneous group. Secondly, the methodology introduces a novel application of the TVP-VAR method combined with Fourier transform to explore the connectedness between cryptocurrency and green financial market. This approach grounded in the DY spillover index, enables an examination of spillover effects across various frequencies, encompassing price movements, returns, and volatility. Furthermore, we also utilize the DCC-GARCH and Copula-GARCH models to delve deeper into the portfolio diversification between these two distinct markets. The insights gained from our analysis can inform more targeted portfolio diversification strategies. Thirdly, we extend the existing knowledge by exploring not only the dynamic connectedness of return series but also that of volatility series. This additional focus allows us to delve into risk spillovers with greater precision.

The remainder of our paper is structured as follows. A review of some relevant prior studies is provided in Section 2. The TVP-VAR and Multivariate GARCH models used in our paper are introduced in Section 3. Section 4 presents some baseline analysis. The time-varying spillover results are presented in Section 5. Section 6 includes the results of hedging performance analysis. Section 7 and Section 8 discusses the results and concludes the paper.

## 2. Literature review

In this section, we discuss the dynamic linkages and hedging performance between cryptocurrencies and the stock market, in particular the sustainable cryptocurrencies. We focus on the recent literature corresponding to the connectedness of cryptocurrencies and the green financial market.

### 2.1 *Spillover effects and hedging performance of cryptocurrencies*

There is a growing scholarly focus on the dynamic connections between cryptocurrencies and the stock market for risk diversification and hedging. Guesmi et al. (2019) show that cryptocurrencies exhibit high average returns and display a weak correlation with traditional financial assets. Moreover, according to Gil-Alana, Abakah and Rojo (2020), there is no evidence to show the cointegration between cryptocurrencies and the stock market indices which suggesting that cryptocurrencies may be a diversification option for investors. Vidal-Tomás and Ibañez (2018) also observed that the self-correlation of Bitcoin has become increasingly effective over time, only influenced by monetary policy news. The finding highlights the difficulty of control over Bitcoin. This finding can provide targeted investment advice for investors with different preferences. Previous studies have constructed connectedness indexes (Diebold and Yilmaz, 2009; Diebold and Yilmaz, 2012; Antonakakis and Gabauer 2017), which have become an effective tool for the measurement of spillover effects of cryptocurrency. The above research lays the foundation for our study, which is that adding cryptocurrencies to an investment portfolio can provide investors with the opportunity to diversify their returns. Besides, on the premise that cryptocurrencies are weakly correlated with other financial assets, utilizing cryptocurrencies for hedging also provides an option for investors seeking risk aversion. Based on the above model, our study extends the underlying TVP-VAR model to a TVP-VAR model with Fourier Transform to investigate the correlation between cryptocurrencies and green financial markets at dynamic frequencies (short, medium and long term).

Moreover, previous research has also confirmed that the unique properties of the dynamic interactions between cryptocurrencies and stock markets make cryptocurrencies effective for hedging. Kajtazi and Moro (2019) add Bitcoin to an optimal portfolio of U.S. and European financial assets, and the results show that the portfolio performance improves. Dwita Mariana et al. (2021) found that both Bitcoin and Ether serve as a short-term haven for stocks, and that Ether may be a better tool for risk avoidance. The hedging performance of cryptocurrencies also verified in emerging financial markets. For instance, Wang et al. (2019) concentrated on the Chinese market and presented that Bitcoin can serve as a hedge against stocks, bonds, and Shanghai Interbank Offered Rate (SHIBOR). It is an effective hedging asset during extreme price changes in the currency markets. Another example is that Bitcoin shows strong safe haven characteristics effectively hedging against the fluctuations of the Malaysia stock index during the COVID-19 period (Ustaoglu, 2022). Previous studies have effectively argued the effectiveness of cryptocurrencies in risk hedging, but mainly focus on the characteristics of traditional cryptocurrencies, such as Bitcoin, and there is still a large gap in studying the hedging performance of sustainable cryptocurrencies that are more environmentally friendly. Our study aims to fill this gap and will concentrate on the hedging effectiveness of sustainable cryptocurrencies to explore whether they can be a green alternative to traditional cryptocurrencies in investment portfolios.

### *2.2 Responsible investment and sustainable cryptocurrencies*

With the rapid growth of cryptocurrencies, there tends to be a greater focus on the conflict between the process of mining cryptocurrencies and environmental protection. Therefore, investors have shifted their focus to hold green financial instruments, including investing in sustainable cryptocurrencies (Fahmy, 2022). Ren and Lucey (2022) split cryptocurrencies into two categories, called “dirty” and “clean” cryptocurrencies, based on their energy consumption levels. In addition to the environmentally friendly characteristics, green cryptocurrencies exhibit the same decentralisation, anonymity and ease of transaction features as traditional cryptocurrencies. Due to their independence from traditional financial institutions, green cryptocurrencies have a reduced correlation with the traditional financial market. Consequently, in periods of fluctuating market conditions, the price fluctuations of green cryptocurrencies may exhibit divergent trends from those of traditional financial assets. Ali et al. (2024) note some similarities between sustainable cryptocurrencies and traditional cryptocurrencies in portfolio management, indicating sustainable cryptocurrencies can be used as an alternative to traditional cryptocurrencies in investment strategies. Similar conclusions can be found in Jacquet and Mans (2019).

Meanwhile, green financial markets have been getting wide attention from investors, because of the returns on investment from these emerging industries and the contribution that the investments can make on sustainable socio-economic development. Green financial markets are correlated with traditional financial markets in terms of asset price movements. For instance, green bonds, due to the debt nature, offer relatively stable returns and lower risks. The returns of green bonds mainly from interest and principal repayment, are closely linked to market interest rates and credit risks, providing some hedging in portfolios. Green stocks, as equity assets, see prices affected by corporate performance, industry prospects, and macroeconomic conditions, making them more sensitive to economic cycles and market environments (Tiwari et al., 2023). Reboredo (2018) found that, the diversification benefits for stock and energy markets offered by green bonds are considerable. Naeem et al. (2023) built a static dependency network between cryptocurrencies and the alternative energy markets to assess their centrality. Ahmed et al. (2024) investigated the spillover effect in the carbon-energy system under the occurrence of extreme events.

The performance of bonds, stocks, and cryptocurrencies differ across economic cycle stages and market shocks (Bashir et al, 2016; Bashir et al, 2019), which provides the possibility for risk diversification. During economic booms, green stocks may yield better returns from corporate profit growth, while green bond returns are relatively stable. In recessions, green bonds may have stronger downside protection, whereas green stocks may face greater downside risks (Corbet et al., 2020). Green cryptocurrencies could display distinct price movements from traditional assets during high market volatility, due to investors' safe-haven demands or speculative activities, thus serving a hedging role in investment portfolios (Li and Meng 2022; Le, 2023). Taking the outbreak of COVID-19 as an example, according to Huang et al. (2023), in the aftermath of the outbreak, green assets consistently served as an effective hedge for Bitcoin. In addition, Karim et al. (2024) studied the extreme risk spillovers which revealed that COVID-19 transformed the spillovers between green bonds and financial markets except Bitcoin. These studies offer a promising foundation for further examining the dynamic linkages and portfolio diversification opportunities between cryptocurrencies and the green financial market during periods of market turbulence. Furthermore, the conclusions drawn from our research can help investors effectively hedge their risk in the event of an extreme situation.

In summary, previous studies have employed different methods to examine the dynamic relationship between cryptocurrency and financial markets, considering the feasibility for the hedging role of cryptocurrency in financial market. However, the past studies concerning cryptocurrency, especially the correlation between sustainable cryptocurrency and the green financial market are very scarce, and the risk-hedging efficiency of sustainable cryptocurrencies in particular has yet to be proven. In addition to this, it is an important task to utilize sustainable cryptocurrencies to assist investors in hedging their risks in the current context of increased global uncertainty. In that case, this paper attempts to fill this gap and construct optimal portfolios between cryptocurrency and the green financial market. Moreover, this paper also considers the differences between conventional cryptocurrencies and sustainable cryptocurrencies, in order to explore the possibility of sustainable cryptocurrencies in portfolio diversification. We use methods including TVP-VAR model with Fourier transform and multivariate GARCH models to address the issue of time-varying return and volatility connectedness between the two markets.

# 3. Methodology

## *3.1 Time-varying parameter vector autoregression (TVP-VAR)*

The Time-Varying Parameter Vector Autoregression (TVP-VAR) model is an effective method of statistical analysis related to financial markets or cryptocurrencies (Youssef et al., 2021; Cui and Maghyereh, 2022; Zeng et al. 2025b). Antonakakis and Gabauer (2017) introduced this model as an extension of the vector autoregressive (VAR) model that captures the dynamic characteristics of model parameters over time. The connectedness obtained by the TVP-VAR model reacts instantly to events, which is particularly suitable for financial markets subject to structural changes and external shocks. But the preceding rolling window VAR smooth out the effect or is sensitive to severe outliers (Antonakakis et al., 2020). By using the Generalized Forecast Error Variance

Decomposition, the *total connectedness index*, which shows how shocks to one variable spill over into other variables, can be derived as follows:

$$C_t^g(J) = \frac{\sum_{i,j=1,i\neq j}^{N} \tilde{\phi}_{ij,t}^{g}(J)}{N} \times 100 \tag{1}$$

where $\tilde{\phi}_{ij,t}^{g}(J)$ represents the proportion of forecast error variance attributed to variable $j$ in explaining the variance of variable $i$ at time $t$ for group $g$ with respect to the forecast horizon $J$. $N$ is the total number of variables in the system. The summation $\sum_{i,j=1,i\neq j}^{N} \tilde{\phi}_{ij,t}^{g}(J)$ sums the forecast error variance spillover across all pairs of variables (excluding the diagonal, where $i = j$).

Regarding the total directional connectedness to others, we initially examine the scenario where variable $i$ imparts its shock to all other variables, denoted as $j$:

$$C_{i\to j,t}^{g}(J) = \frac{\sum_{j=1,i\neq j}^{N} \tilde{\phi}_{ji,t}^{g}(J)}{\sum_{j=1}^{N} \tilde{\phi}_{ji,t}^{g}(J)} \times 100. \tag{2}$$

Furthermore, we compute the directed connectedness received by variable $i$ from variable $j$, called the total directional connectedness from others. This index is defined as follows:

$$C_{i\leftarrow j,t}^{g}(J) = \frac{\sum_{j=1,i\neq j}^{N} \tilde{\phi}_{ij,t}^{g}(J)}{\sum_{i=1}^{N} \tilde{\phi}_{ij,t}^{g}(J)} \times 100. \tag{3}$$

Finally, according to the Equation (2) and (3), the *net total directional connectedness* is derived. This difference can be interpreted as the impact of variable $i$ on the entire network of variables. A positive *net total directional connectedness* for variable $i$ suggests that it exerts more influence on the network than it receives from the network[6]:

$$C_{i,t}^{g} = C_{i\to j,t}^{g}(J) - C_{i\leftarrow j,t}^{g}(J). \tag{4}$$

[6] For more detailed model derivations and explanations of Section 3.1, please refer to the appendix.

## 3.2 *The connectedness network based on frequency decomposition*

In order to better represent the connectedness of information spillovers, we extend our study from the connectedness of a single given frequency to the connectedness of dynamic frequencies (short-term, medium-term and long-term). According to Baruník and Křehlík (2018), a frequency band is defined as the amount of forecast error variance generated over a set of convex frequencies. The generalized variance decompositions on frequency band $d$, where $d = (a, b): a, b \in (-\pi, \pi), a < b$, are defined as:

$$\theta_{i,j}(d) = \frac{1}{2\pi} \int_d \Gamma_i(\omega) \mathbf{f}(\omega)_{i,j} \, \mathrm{d}\omega \tag{5}$$

where the frequency variable $\omega \in (-\pi, \pi)$. $\Gamma_i(\omega)$ is the weighting function and $\mathbf{f}(\omega)_{i,j}$ is the generalized causation spectrum.

The within connectedness reflects the effect of connectedness within a frequency band and is derived by weighting the power of sequences on a particular frequency band. The equation of within connectedness is

$$\mathrm{S}^{\mathrm{M}}(\mathrm{d}) = 100 \times (1 - \frac{\mathrm{Tr}\{\hat{\theta}(d)\}}{\Sigma\hat{\theta}(d)}) \tag{6}$$

where $\mathrm{Tr}\{\cdot\}$ is the trace operator, and $\hat{\theta}(d)$ is the scaled generalized variance decomposition. The frequency connectedness is defined as[7]:

$$\mathrm{S}^{\mathrm{F}}(\mathrm{d}) = \mathrm{S}^{\mathrm{M}}(\mathrm{d}) \times \frac{\Sigma\hat{\theta}(d)}{\Sigma\hat{\theta}(\infty)}. \tag{7}$$

## 3.3 *Multivariate GARCH model*

[7] For more detailed model derivations and explanations of Section 3.2, please refer to the appendix.

### 3.3.1 DCC GARCH model

The dynamic conditional correlation (DCC) model is a time series model commonly used in financial econometrics. This model is particularly useful for analyzing and forecasting the volatility and correlation of multivariate time series data (e.g. financial asset returns). Since this model was first proposed by Engle (2002), it has been used to estimate conditional variance matrices for time-series data and is particularly useful in analyzing the covariance of financial asset returns. There are some advantages of this model. On the one hand, the DCC model can capture and describe the dynamics of correlations between multivariate time series (e.g., returns on multiple financial assets). This is achieved by modelling the time-varying properties of the conditional correlation matrix. On the other hand, the low number of model parameters makes the DCC model highly flexible in practical applications. It can be applied to data of different time spans (e.g., daily, weekly, monthly, etc.).

In this paper, we explore the dynamic relationship between cryptocurrencies and the green financial market, an ARMA-DCC-GJR-GARCH model has been utilized. The first step is to construct an ARMA model as follows:

$$r_t = c + \sum_{i=1}^{m} \phi_i y_{t-i} + \sum_{i=1}^{n} \theta_i \varepsilon_{t-i} + \varepsilon_t \tag{8}$$

where $r_t$ represents the return series calculated using a logarithmic approach. The parameters $\phi_i$ correspond to the autoregressive (AR) component, while $\theta_i$ are the parameters associated with the moving average (MA) component. $\varepsilon_{t-i}$ denotes the residual term and the variables $m$ and $n$ represent the lag orders of the AR and MA processes, respectively.

Concerning the GJR-GARCH model, it serves as an extension of the GARCH (Generalized Autoregressive Conditional Heteroskedasticity) model (Glosten et al., 1993). The GJR-GARCH model captures not only volatility aggregation (i.e., large price fluctuations are often preceded by even larger price fluctuations), but also the so-called leverage effect, whereby negative information or a down market generally leads to a greater increase in volatility than positive information or an up market. Its general form is:

$$\sigma_t^2 = \omega + \sum_{i=1}^{k} (\alpha_i \varepsilon_{t-i}^2 + \gamma_i \varepsilon_{t-i}^2 I_{[\varepsilon_{t-i}<0]}) + \sum_{j=1}^{s} \beta_j \sigma_{t-j}^2 \quad (9)$$

where $\sigma_t^2$ represents the conditional variance, $\omega$, $\alpha_i$, $\gamma_i$ and $\beta_j$ denote the parameters of this model. $I_{[\varepsilon_{t-i}<0]}$ is the indicator function used to capture the leverage effect, taking the value of 1 when $\varepsilon_{t-i} < 0$ and 0 otherwise. Moreover, $k$ and $s$ are the lag orders of the GARCH and ARCH terms.

For DCC model, the core of this model is to describe changes in correlation between time series. The fundamental DCC system can be represented by the following expressions:

$$Q_t = (1 - \sum_{i=1}^{a} \alpha_i - \sum_{j=1}^{b} \beta_j)\overline{Q} + \sum_{i=1}^{a} \alpha_i \varepsilon_{t-i} \varepsilon_{t-i}' + \sum_{j=1}^{b} \beta_j Q_{t-j} \quad (10)$$

$$R_t = \mathrm{diag}(Q_t)^{-1/2} Q_t \mathrm{diag}(Q_t)^{-1/2} \quad (11)$$

where $Q_t$ is an unstandardised version of the correlation matrix over time and $\overline{Q}$ is the long-term average correlation matrix. $\alpha_i$ and $\beta_j$ are parameters capturing the speed of adjustment of the dynamic correlations. $R_t$ is the normalised conditional correlation matrix. $a$ and $b$ are the orders of the model parameters.

Combining these models, we first use an ARMA model to simulate the mean equation of the time series, then a GJR-GARCH model to capture asymmetries in volatility, and finally, a DCC model to capture dynamic correlations between multiple time series. This integrated approach is very effective in dealing with the complexity of financial markets as it allows multiple time series properties to be considered simultaneously.

### 3.3.2 Copula GARCH model

Copula is a mathematical tool for describing the dependence structure of random variables. It separates the marginal distributions of multidimensional distributions from the correlation structure, allowing the modelling of independence and correlation between the two. In finance, Copula models are often used to model correlations between different assets, especially in the case of extreme events (Nelsen, 2006). The probability distribution function of binary normal Copula is shown below:

$$C_G(u, v; \rho) = \frac{1}{\sqrt{1-\rho^2}} exp\{-\frac{\Phi^{-1}(u)^2 + \Phi^{-1}(v)^2 - 2\rho\Phi^{-1}(u)\Phi^{-1}(v)}{2(1-\rho^2)}\} \tag{12}$$

where $u$ and $\rho$ are two time series data of cryptocurrencies and green financial indices. Similar to the DCC model described above, these values are estimated from GJR-GARCH models. $\Phi(\cdot)$ represents the normal distribution function of the time series, and $\rho \in (-1,1)$ is the correlation coefficient of $\Phi^{-1}(u)$ and $\Phi^{-1}(v)$.

In our paper, we use a combination of Copula and GARCH models to model dependencies and volatility in financial time series. The advantage of this model is that it can capture the correlation between different assets more accurately and take into account changes in volatility in the time series. Unlike traditional linear models such as Granger causality or static correlation analysis, the copula framework allows for modeling asymmetric dependence and tail co-movements, which are especially relevant in high-volatility environments. Compared to the DCC model, the Copula model is more flexible for modelling correlations that deal with non-normal marginal distributions and extreme events.

### *3.4 Hedging performance*

The results obtained from estimating the DCC model and Copula model can be utilized to formulate portfolios, determining two approaches to evaluate the constructed portfolios at various stages. This procedure aids in offering investment recommendations for investors.

Initially, when addressing the challenge of estimating a dynamically risk-minimizing hedge ratio using multivariate GARCH models, we adopt the beta hedge method presented by Kroner and Ng (1998). The risk-minimizing hedge ratio β is determined by the following equation:

$$\beta_t^* = \frac{h_{12,t}}{h_{22,t}} \tag{13}$$

where $h_{12,t}$ is the covariance between the returns of two assets (asset 1 and asset 2) at time $t$, estimated using a multivariate GARCH model. $h_{22,t}$ is the variance of the returns of asset 2 at time $t$.

Secondly, employing the Hedging Effectiveness (HE) index is an important step we use to evaluate how well our ideal portfolios hedge (Jiang et al., 2019). A higher HE represents a better hedging performance. The fundamental expression is shown below:

$$HE = 1 - \frac{Var_p}{Var_0} \tag{14}$$

where $Var_p$ represents the variance of our optimal portfolios and $Var_0$ represents the variance of benchmark portfolio.[8]

# 4. Baseline analysis

## *4.1 Data*

The objective of this study is to explore the dynamic connectedness of return and volatility series in cryptocurrencies and the green financial markets and investigate the opportunities of portfolio diversification between the two markets. We also pay attention to the impact of market shocks, such as the COVID-19 and the Russia-Ukraine war. The COVID-19 pandemic caused unprecedented global economic disruptions, which led to heightened market volatility and changes in correlations between assets, particularly in the cryptocurrency and green financial markets (Li and Yan, 2022). Similarly, the Russia-Ukraine war is considered a major external shock that has influenced global financial markets by amplifying risk aversion, increasing energy prices, and disrupting supply chains (Umar et al., 2022). These events provide a natural setting to examine the impact of extreme market shocks on the interconnectedness of assets in the study, and the conclusions drawn from our research can help investors effectively hedge their risk in the event of an extreme situation.

The first step is to collect the cryptocurrency data. Bitcoin (BTC) is adopted as the representative of conventional cryptocurrency. In addition, Cardano (ADA), Ripple (XRP), IOTA (MIOTA), and Stellar (XLM) are four of the most traded green cryptocurrencies, respectively adopting the Proof of Stake (PoS) mechanism that randomly selects validators, the Ripple Protocol Consensus

[8] All models in this paper are programmed and implemented based on R language.

Algorithm (RPCA) that reaches consensus through a group of validation nodes, the Tangle technology that reaches consensus through voting among nodes, and the Stellar Consensus Protocol (SCP) mechanism based on Directed Acyclic Graph (DAG). These consensus mechanisms do not rely on a large amount of computational work and are more environmentally friendly, so they are emphasized in the analysis. Green energy cryptocurrencies represented by Powerledger (POWR) and SunContract (SNC) achieve the trading and management of energy through blockchain technology, promoting the decentralization and sustainable development of the energy industry. For instance, solar power generation users can sell their surplus electricity to other users via Powerledger or SunContract without going through traditional energy suppliers. The daily prices of target cryptocurrencies are collected from coinmarketcap.com[9].

Regarding the data of green financial markets, we utilize three categories of indices referring to Zeng et al. (2025a), Reboredo and Ugolini (2020), and Karim et al. (2024). For the bond market, we use the S&P Green Bond Index (GBI) as it tracks global green bonds financing environmentally sustainable projects. For the stock market, we select the S&P Global 1200 ESG Index (ESG) and the S&P Global Clean Energy Index (CEI). ESG represents firms with strong environmental, social, and governance practices, while CEI focuses on clean energy companies, together capturing key aspects of sustainability in equities. For the sustainable energy market, the Global Wind Energy Index (WIND) and the World Solar Energy Index (SOLAR) represent the financial dynamics of wind and solar energy sectors. The WIND index reflects the performance of companies engaged in wind energy production, while the SOLAR index tracks firms specializing in solar energy solutions, providing a direct measure of the financial viability and investor sentiment within the renewable energy sector. The data is collected from Refinitiv.

[9] https://coinmarketcap.com

In particular, the time series covers the period 2018.1.2 to 2025.02.06, with daily observations. The selection of the starting point is primarily driven by data availability and market maturity considerations. Prior to 2018, the green financial market data, especially for daily traded indices were relatively scarce or inconsistent. Moreover, starting from 2018 ensures that both the cryptocurrency market and the green financial market had reached a level of liquidity and structure suitable for rigorous empirical analysis, thereby enhancing the reliability of the results. Referring to Bouteska et al. (2023), the entire dataset can be split into three segments: before COVID-19 period (2018.1.2 to 2019.12.31), during COVID-19 period (2020.1.1 to 2023.02.23) and during Russia-Ukraine war (2023.02.24 to 2025.02.06) to measure the effect of market shock. Additionally, in order to handle data discrepancies between cryptocurrency markets (which are traded 24/7) and green financial markets (which are closed on holidays), we manually remove the data from the days when they are not traded.

### *4.2 Descriptive statistics*

For a direct visualization of data, the time trends of daily price series are depicted in Figure 1. We use the red dotted line to distinguish between periods. From the picture, it can be seen that the price of conventional cryptocurrency shows a different trend from the sustainable cryptocurrencies. BTC shows increasing trends after the market shocks while the price of sustainable cryptocurrencies still maintains a low level after the occurrence of Russia-Ukraine war. BTC also shows higher volatility with sharp price swings, reflecting its speculative nature, while sustainable cryptocurrencies (XRP,

ADA, XLM, MIOTA, POWR, SNC) exhibit relatively lower fluctuations, possibly due to distinct investor sentiment or institutional backing. The red dotted lines indicate major volatility events, which impact BTC more significantly, suggesting its greater sensitivity to macroeconomic shocks and regulatory changes. In contrast, sustainable cryptocurrencies appear more stable, potentially due to their niche utility and long-term environmental focus. Regarding the green financial indices, despite the negative effects of COVID-19, the prices initially fall but soon exhibit an upward trend due to quantitative easing, indicating that people are still willing to invest in green financial assets. Moreover, Russia-Ukraine war leads to higher energy prices and exacerbate environmental problems and further accelerates the demand for green energy. This may prompt investors to be more inclined to invest in renewable energy projects in order to reduce their dependence on conventional energy sources. The continued rise in the price of ESG and CEI confirms this viewpoint.

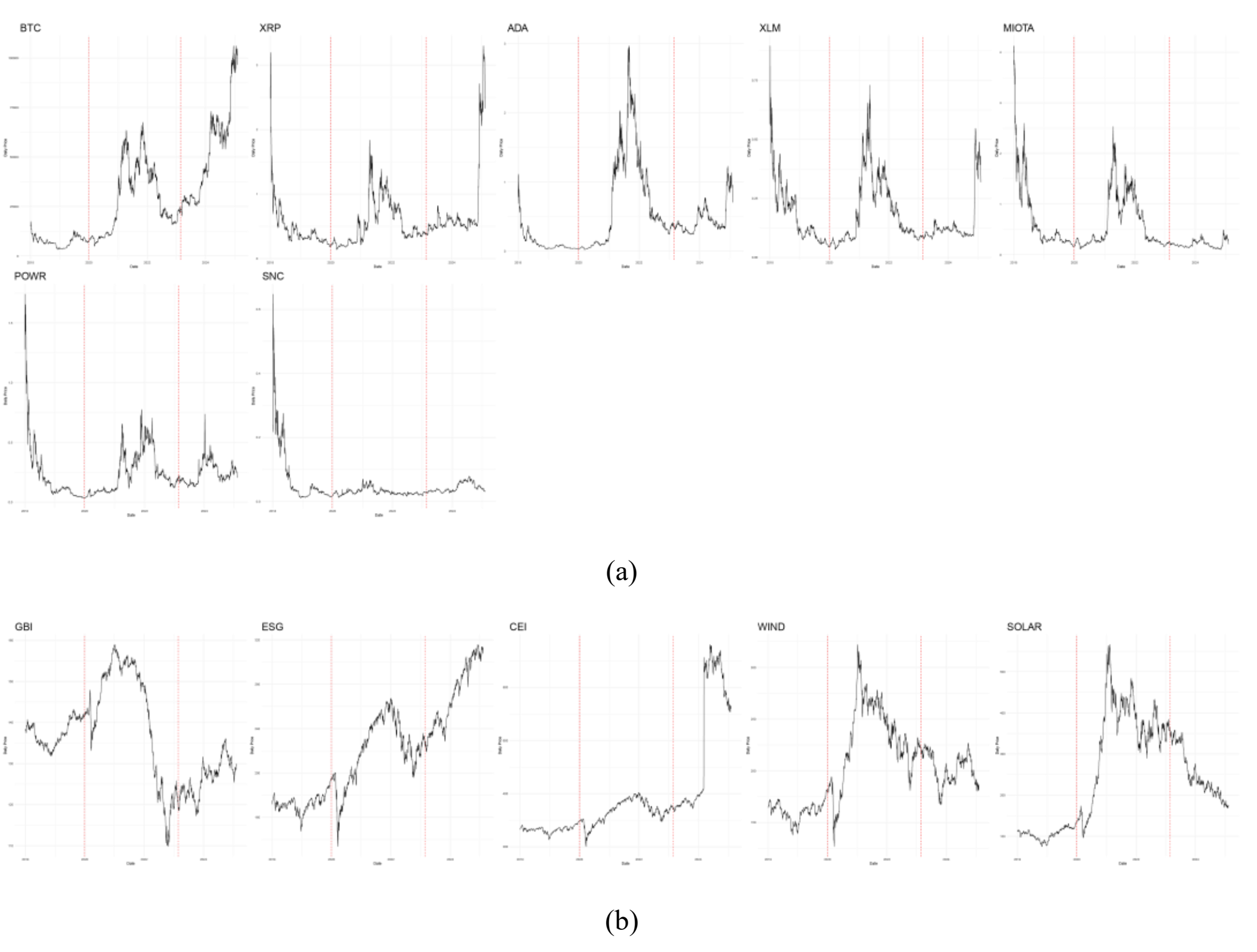

Figure 1. Daily price trends of (a) traditional cryptocurrency (BTC), sustainable cryptocurrencies (XRP, ADA, XLM,

MIOTA, POWR, SNC), and (b) green financial indices (GBI, CEI, ESG, WIND, SOLAR). The red dotted lines mark significant volatility events.

The statistical results for the daily log-return series of selected cryptocurrencies and green financial indices are presented in Table 1. During the pre-COVID-19 period, most cryptocurrencies exhibit negative average returns, while their mean returns become positive during and after the onset of COVID-19, indicating a potential shift in market dynamics and growing investor interest. In contrast, green financial indices generally maintain positive average returns throughout the three periods, except for WIND and SOLAR, which show slightly negative means in certain phases. Standard deviations for cryptocurrencies are consistently higher than those of green financial indices, underscoring the elevated volatility and speculative nature of crypto assets. Notably, POWR exhibits the highest volatility across all periods. Regarding normality, the Jarque-Bera (J-B) test rejects the null hypothesis at the 1% level for all return series, indicating significant deviations from normal distribution. Together with the results of the ARCH-LM test in Table 2, this provides strong evidence of heavy tails and time-varying volatility, thereby justifying the application of GARCH-family models (Engle, 1982; Bollerslev, 1986; Hansen, 1994). To assess stationarity, we employed both the Augmented Dickey-Fuller (ADF) test and the Kwiatkowski-Phillips-Schmidt-Shin (KPSS) test. The ADF test results reject the null hypothesis of unit root, while the KPSS test fails to reject the null hypothesis of stationarity for all series, jointly confirming that the return series are stationary. Stationarity is a prerequisite for applying autoregressive moving-average (ARMA) models, which, combined with the evidence of conditional heteroskedasticity supports the use of the ARMA-GJR-GARCH framework in this study (Hamilton JD, 2020).

Table 1: Statistical results of daily log-return series

| | Mean | Max | Min | S.D. | Skew | Kurt | J-B | ADF | KPSS |
|---|---|---|---|---|---|---|---|---|---|
| **Pre_COVID-19** | | | | | | | | | |
| BTC | -0.0014 | 0.2030 | -0.2387 | 0.0457 | -0.3023 | 3.5389 | 281.0455*** | -7.5614*** | 0.2652 |
| XRP | -0.0050 | 0.3220 | -0.3520 | 0.0631 | 0.1427 | 4.8410 | 511.9730*** | -8.4537*** | 0.1954 |
| ADA | -0.0061 | 0.3218 | -0.2173 | 0.0677 | 0.3860 | 2.4924 | 148.7519*** | -7.4179*** | 0.2325 |
| XLM | -0.0049 | 0.4618 | -0.3062 | 0.0669 | 0.6079 | 5.9816 | 810.2348*** | -8.3502*** | 0.0327 |
| MIOTA | -0.0062 | 0.2766 | -0.2918 | 0.0689 | 0.0179 | 2.3624 | 122.1410*** | -8.0361*** | 0.1380 |
| POWR | -0.0070 | 0.3485 | -0.3862 | 0.0927 | -0.1017 | 3.8381 | 322.0164*** | -8.4871*** | 0.1149 |
| SNC | -0.0051 | 0.8970 | -0.4742 | 0.0738 | 2.2564 | 23.0934 | 12007.3996*** | -7.9846*** | 0.1521 |
| GBI | 0.0001 | 0.0075 | -0.0074 | 0.0021 | 0.0600 | 0.3304 | 2.8468*** | -8.4351*** | 0.2968 |
| ESG | 0.0002 | 0.0266 | -0.0311 | 0.0072 | -0.6701 | 2.1276 | 137.9968*** | -7.9735*** | 0.1458 |
| CEI | 0.0002 | 0.0273 | -0.0312 | 0.0072 | -0.6616 | 2.0916 | 133.7073*** | -7.9488*** | 0.1374 |
| WIND | 0.0002 | 0.0313 | -0.0262 | 0.0074 | -0.1429 | 0.9425 | 21.4893*** | -7.8522*** | 0.2176 |
| SOLAR | 0.0003 | 0.0502 | -0.0475 | 0.0144 | -0.1195 | 0.3045 | 3.4014*** | -8.8771*** | 0.2388 |
| **During_COVID-19** | | | | | | | | | |
| BTC | 0.0015 | 0.1915 | -0.4647 | 0.0454 | -1.6057 | 15.6922 | 8783.9804*** | -8.6502*** | 0.3994 |
| XRP | 0.0009 | 0.6267 | -0.5505 | 0.0722 | 0.1775 | 15.6723 | 8414.5335*** | -9.3080*** | 0.1248 |
| ADA | 0.0030 | 0.2794 | -0.5037 | 0.0688 | -0.2211 | 5.4219 | 1015.0347*** | -8.4585*** | 0.6849 |
| XLM | 0.0009 | 0.5592 | -0.4100 | 0.0674 | 0.6053 | 11.7802 | 4803.2319*** | -10.4940*** | 0.3336 |
| MIOTA | 0.0006 | 0.3321 | -0.5436 | 0.0723 | -0.8736 | 9.0215 | 2892.9876*** | -9.2956*** | 0.3090 |
| POWR | 0.0022 | 0.4717 | -0.7003 | 0.0838 | -0.7519 | 11.3206 | 4466.9869*** | -9.8097*** | 0.2004 |
| SNC | 0.0007 | 0.6093 | -0.4986 | 0.0766 | -0.1247 | 10.5594 | 3821.7289*** | -9.9609*** | 0.0547 |
| GBI | -0.0002 | 0.0227 | -0.0241 | 0.0046 | -0.2477 | 3.8064 | 505.9297*** | -8.5352*** | 0.4234 |
| ESG | 0.0002 | 0.0821 | -0.1022 | 0.0130 | -0.9855 | 12.3819 | 5383.5446*** | -8.2345*** | 0.1051 |
| CEI | 0.0002 | 0.0854 | -0.1045 | 0.0132 | -1.0089 | 12.8929 | 5832.0272*** | -8.2453*** | 0.0968 |
| WIND | 0.0002 | 0.0989 | -0.1259 | 0.0151 | -0.6657 | 9.3664 | 3066.3967*** | -8.3715*** | 0.1820 |
| SOLAR | 0.0012 | 0.1064 | -0.1466 | 0.0242 | -0.4057 | 3.6720 | 485.6050*** | -8.1302*** | 0.3771 |
| **During_R-U war** | | | | | | | | | |
| BTC | 0.0028 | 0.1812 | -0.1288 | 0.0318 | 0.7774 | 3.8275 | 361.6707*** | -8.0030*** | 0.0517 |
| XRP | 0.0036 | 0.5486 | -0.1485 | 0.0533 | 3.4741 | 29.1016 | 18886.7318*** | -6.5310*** | 0.2213 |
| ADA | 0.0012 | 0.3243 | -0.2004 | 0.0487 | 0.8256 | 5.7974 | 768.9235*** | -6.3519*** | 0.0984 |
| XLM | 0.0025 | 0.4756 | -0.1853 | 0.0542 | 3.5016 | 26.1094 | 15418.2760*** | -6.2690*** | 0.1355 |
| MIOTA | -0.0004 | 0.3884 | -0.2227 | 0.0587 | 1.2300 | 8.1495 | 1531.7390*** | -6.9822*** | 0.0659 |
| POWR | 0.0000 | 0.4620 | -0.4952 | 0.0651 | -0.0576 | 13.8739 | 4065.6505*** | -8.6107*** | 0.0744 |
| SNC | 0.0001 | 0.1336 | -0.2299 | 0.0427 | -0.3353 | 3.1316 | 217.8077*** | -7.6749*** | 0.2464 |
| GBI | 0.0002 | 0.0192 | -0.0113 | 0.0042 | 0.5053 | 1.5829 | 75.1204*** | -7.0624*** | 0.0715 |
| ESG | 0.0006 | 0.0278 | -0.0347 | 0.0068 | -0.3181 | 1.7585 | 74.6026*** | -8.3214*** | 0.0342 |
| CEI | 0.0015 | 0.7793 | -0.0627 | 0.0361 | 19.8534 | 424.8280 | 3838634.977*** | -7.7514*** | 0.1232 |
| WIND | -0.0003 | 0.0662 | -0.0535 | 0.0112 | 0.1649 | 4.6712 | 464.7089*** | -7.6774*** | 0.0886 |
| SOLAR | -0.0014 | 0.0690 | -0.0935 | 0.0173 | 0.0081 | 2.1361 | 97.2927*** | -8.4850*** | 0.0272 |

These preliminary findings—high volatility, non-normality, and stationarity—justify our use of volatility modeling frameworks and motivate the investigation of dynamic connectedness and diversification potential between the cryptocurrency and green financial markets in the presence of extreme shocks.

Figure 2 presents a heatmap of the pairwise Pearson correlation among cryptocurrency returns and green financial indices. The results indicate that all correlation values between the twelve assets are positive, though the strength of these correlations varies. Specifically, the connections between cryptocurrencies and green financial assets are weak by distinct market forces. This weak correlation implies that incorporating sustainable cryptocurrencies into a portfolio alongside green financial assets may enhance diversification benefits by reducing overall risk exposure. Notably, ESG shows a relatively high correlation with WIND and SOLAR, indicating a stronger connection between ESG-focused stocks and the renewable energy sector. Furthermore, WIND and SOLAR exhibit the highest correlation, reflecting the close relationship between these two renewable energy markets. These patterns highlight the strong internal linkages within the green financial sector, suggesting that renewable energy stocks and ESG investments may move in tandem. Additionally, a high correlation is observed among cryptocurrencies, supporting the idea that sustainable cryptocurrencies can serve as an alternative to conventional cryptocurrencies in portfolio construction, consistent with the view proposed by Ali et al. (2024).

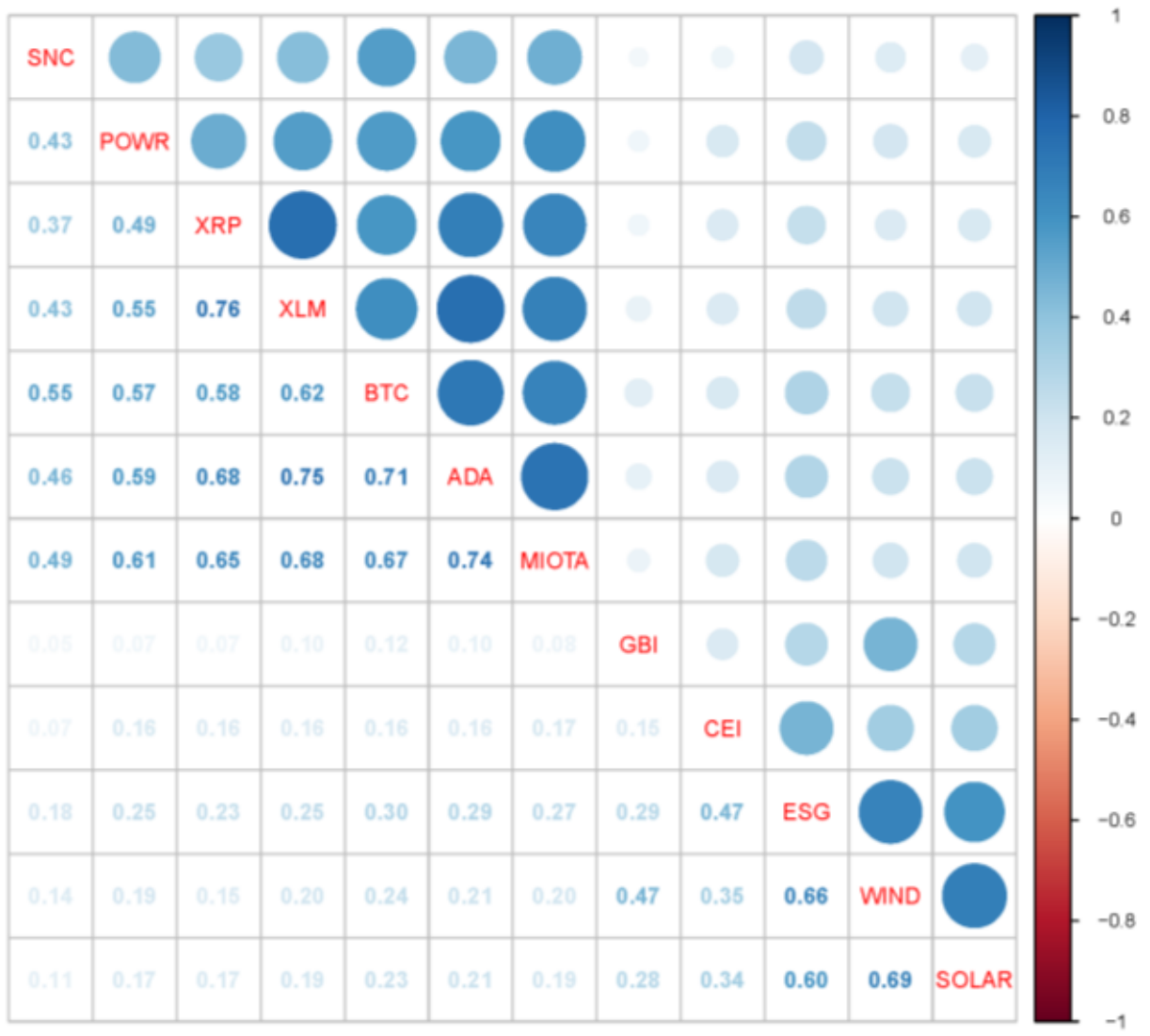

Figure 2. Correlation heat map

Based on the summary statistics of the log-return series, it is appropriate to preprocess the series by GARCH model to estimate the conditional heteroskedastic volatility, because the model allows the variance to change across different periods and enhances the accuracy of fitting and forecasting. Specifically, we adopt the Glosten-Jagannathan-Runkle GARCH (GJR-GARCH) model, which is capable of capturing the leverage effect induced by positive or negative market shocks. Before that, the conditional means of time series data are estimated by appropriate ARMA model, whose selection is based on the smallest Akaike information criterion (AIC). Table 2 presents the Ljung-Box and LM-test results along with the optimal ARMA and GJR-GARCH models. According to the Ljung-Box results, the residual series exhibit no autocorrelation, indicating that the residuals are white noise, and the model's parameter estimates are valid. The LM-test results show that the ARCH effect is significant at a 1% significance level. Notably, a normal distribution is used to fit the standardized residuals during model estimation.

Table 2: Valid ARMA-GJRGARCH models

| | Ljung-Box | ARCH LM-test | Mean Model | GARCH Model |
|---|---|---|---|---|
| BTC | 13.4160 | 41.8450$^{***}$ | ARMA(2,4) | GJRGARCH(1,1) (normal distribution) |
| XRP | 8.5012 | 66.7230$^{***}$ | ARMA(3,4) | GJRGARCH(1,1) (normal distribution) |
| ADA | 18.2870 | 87.0790$^{***}$ | ARMA(3,3) | GJRGARCH(1,1) (normal distribution) |
| XLM | 6.6010 | 90.2690$^{***}$ | ARMA(1,5) | GJRGARCH(1,1) (normal distribution) |
| MIOTA | 11.7690 | 106.2500$^{***}$ | ARMA(1,0) | GJRGARCH(1,1) (normal distribution) |
| POWR | 12.9660 | 72.6280$^{***}$ | ARMA(3,3) | GJRGARCH(1,1) (normal distribution) |
| SNC | 18.0430 | 137.9900$^{***}$ | ARMA(3,2) | GJRGARCH(1,1) (normal distribution) |
| GBI | 18.3850 | 239.0900$^{***}$ | ARMA(1,0) | GJRGARCH(1,1) (normal distribution) |
| ESG | 18.7990 | 658.4700$^{***}$ | ARMA(5,5) | GJRGARCH(1,1) (normal distribution) |
| CEI | 21.4690 | 457.2390$^{***}$ | ARMA(5,5) | GJRGARCH(1,1) (normal distribution) |
| WIND | 9.5002 | 409.4900$^{***}$ | ARMA(5,5) | GJRGARCH(1,1) (normal distribution) |
| SOLAR | 26.7250 | 310.6800$^{***}$ | ARMA(4,3) | GJRGARCH(1,1) (normal distribution) |

Note: Ljung-Box test is used to check residual autocorrelation. ARCH LM-test is the ARCH effect test.

Table 3: Optimal parameter estimations on ARMA-GJRGARCH model

| | BTC | XRP | ADA | XLM | MIOTA | POWR | SNC | GBI | ESG | CEI | WIND | SOLAR |
|---|---|---|---|---|---|---|---|---|---|---|---|---|
| $\emptyset_1$ | -0.478$^{***}$ | 0.378$^{***}$ | 1.501$^{***}$ | -0.726$^{***}$ | 0.084$^{***}$ | 0.449$^{***}$ | 0.969$^{***}$ | 0.118$^{***}$ | -0.965$^{***}$ | -0.547$^{***}$ | -1.448$^{***}$ | -0.673$^{***}$ |
| $\emptyset_2$ | -0.577$^{***}$ | 0.163$^{***}$ | -0.661$^{***}$ | \ | \ | -0.362$^{***}$ | -0.903$^{***}$ | \ | -0.183$^{***}$ | 0.814$^{***}$ | -1.341$^{***}$ | 0.926$^{***}$ |
| $\emptyset_3$ | \ | -0.891$^{***}$ | 0.157$^{***}$ | \ | \ | 0.570$^{***}$ | -0.073$^{***}$ | \ | -0.481$^{***}$ | 0.495$^{***}$ | -1.249$^{***}$ | 0.880$^{***}$ |
| $\emptyset_4$ | \ | \ | \ | \ | \ | \ | \ | \ | -0.947$^{***}$ | -0.762$^{***}$ | -0.488$^{***}$ | -0.143$^{***}$ |
| $\emptyset_5$ | \ | \ | \ | \ | \ | \ | \ | \ | -0.383$^{***}$ | -0.774$^{***}$ | 0.119$^{***}$ | \ |
| $\theta_1$ | 0.443$^{***}$ | -0.318$^{***}$ | -1.484$^{***}$ | 0.755$^{***}$ | 0.074$^{***}$ | -0.457$^{***}$ | -1.053$^{***}$ | \ | 1.092$^{***}$ | 0.569$^{***}$ | 1.602$^{***}$ | 0.824$^{***}$ |
| $\theta_2$ | 0.602$^{***}$ | -0.146$^{***}$ | 0.677$^{***}$ | 0.029$^{***}$ | \ | 0.345$^{***}$ | 0.992$^{***}$ | \ | 0.325$^{***}$ | -0.807$^{***}$ | 1.618$^{***}$ | -0.804$^{***}$ |
| $\theta_3$ | 0.039$^{***}$ | 0.850$^{***}$ | -0.195$^{***}$ | -0.091$^{***}$ | \ | -0.616$^{***}$ | \ | \ | 0.503$^{***}$ | -0.528$^{***}$ | 1.522$^{***}$ | -1.003$^{***}$ |
| $\theta_4$ | 0.077$^{***}$ | 0.101$^{***}$ | \ | -0.044$^{***}$ | \ | \ | \ | \ | 1.014$^{***}$ | 0.761$^{***}$ | 0.766$^{***}$ | \ |
| $\theta_5$ | \ | \ | \ | 0.053$^{***}$ | \ | \ | \ | \ | 0.505$^{***}$ | 0.808$^{***}$ | 0.054$^{***}$ | \ |
| $\mu$ | -0.001$^{***}$ | -0.003$^{***}$ | 0.002$^{***}$ | -0.005$^{***}$ | -0.002$^{*}$ | -0.003$^{***}$ | -0.001$^{***}$ | 0.000 | 0.007$^{***}$ | 0.005$^{***}$ | 0.000 | 0.000 |
| $\delta_0$ | 0 | 0 | 0 | 0 | 0 | 0 | 0 | 0 | 0 | 0 | 0 | 0 |
| $\alpha_1$ | 0.009$^{***}$ | 0.018$^{***}$ | 0.050$^{***}$ | 0.135$^{***}$ | 0.074$^{***}$ | 0.056$^{***}$ | 0.006$^{***}$ | 0.045$^{***}$ | 0.001$^{***}$ | 0.004$^{***}$ | 0.038$^{***}$ | 0.054$^{***}$ |
| $\beta_1$ | 0.992$^{***}$ | 0.988$^{***}$ | 0.944$^{***}$ | 0.903$^{***}$ | 0.926$^{***}$ | 0.954$^{***}$ | 0.990$^{***}$ | 0.950$^{***}$ | 0.941$^{***}$ | 0.942$^{***}$ | 0.939$^{***}$ | 0.941$^{***}$ |
| $\gamma_1$ | -0.006$^{***}$ | -0.016$^{***}$ | 0.008 | -0.079$^{***}$ | -0.004 | -0.024$^{***}$ | 0.002 | 0.005 | 0.114$^{***}$ | 0.111$^{***}$ | 0.041$^{***}$ | 0.006 |

Notes: $\emptyset_1$ to $\emptyset_5$ denote the autoregressive parameters. The moving average parameters are marked by $\theta_1$ to $\theta_5$, while $\mu$ is a fixed parameter. The variance equation's constant parameter is denoted by $\delta_0$, the ARCH parameter by $\alpha_1$, the GARCH coefficient by $\beta_1$, and the leverage parameter by $\gamma_1$

Table 3 displays the parameter estimates for the ARMA-GJR-GARCH models. The values of $\emptyset_i$s and $\theta_i$s are statistically significant at 1% level, indicating the mean equations are valid. Moreover, the values of $\alpha_1$ and $\beta_1$ are also statistically significant at 1% level, implying the validity of GJR-GARCH models. Estimated $\gamma_1$ of BTC, XRP, XLM, POWR, ESG, CEI and WIND are statistically significant at 1% level. According to Glosten et al. (1993), in the GJR-GARCH framework, the coefficient $\gamma_1$ captures the asymmetric impact of negative shocks on volatility. The statistical significance of $\gamma_1$ in our results supports the presence of the leverage effect. For green financial indices (ESG, CEI, and WIND), $\gamma_1$ is positive and significant, which confirms the standard leverage effect, where negative shocks such as adverse external news disproportionately increase volatility.

## 5. Measurement of time-varying spillover

In this section, the connectedness analysis of return series using the TVP-VAR approach and Fourier transform are demonstrated, with a specific emphasis on the variations induced by the market shocks. Afterwards, we analyze the portfolio of cryptocurrencies and green financial markets and verify the hedging performance of conventional and sustainable cryptocurrencies.

### *5.1 Connectedness on return series*

This subsection discusses the information spillover of cryptocurrencies and green financial indices. Figure 3 illustrates the dynamic total connectedness of return series, considering the situation at different frequencies. In addition, the connectedness is estimated using the Bootstrap method with a 95% confidence interval, and the shaded areas represent the confidence bounds. Figure 3(a) shows that the dynamic total connectedness of return series fluctuates over time, ranging from 57% to 92%, with notable spikes during major global events. For instance, the peak in early 2020 coincides with the outbreak of the COVID-19 pandemic, and a second rise is observed following the onset of the Russia–Ukraine war. These observations suggest that extreme global events may coincide with intensified information spillovers across markets.

This phenomenon can be partially explained by the increased uncertainty and volatility during crisis periods, as evidenced by wider confidence intervals around the connectedness estimates. As noted by Mun and Brooks (2012), heightened market uncertainty is often associated with stronger cross-market correlations. Moreover, according to the Adaptive Market Hypothesis (Lo, 2004), market efficiency may vary over time, and periods of high volatility tend to impair informational efficiency, which can alter the patterns of information transmission across markets. Therefore, although our results do not establish causality, the timing and dynamics of total connectedness appear consistent with the notion that increased market uncertainty may contribute to enhanced intermarket spillovers.

However, the total connectedness estimated from traditional TVP-VAR does not reveal whether the effect on the system is in short-term or long-term. Investors pay attention to the investment horizon and thus value assets through the expected utility of investment at different levels. These behaviours encourage us to investigate the connectedness via different frequency decomposition.

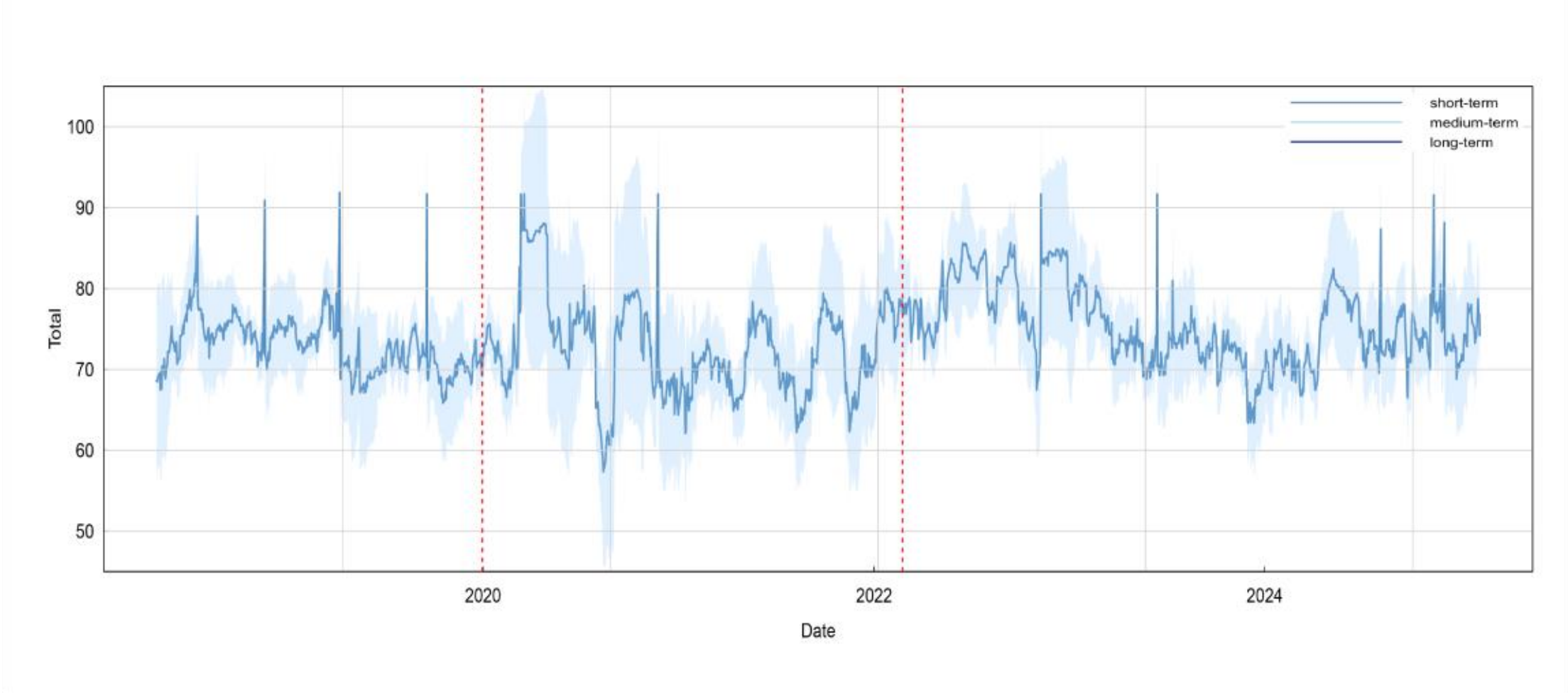

(a) Dynamic total connectedness

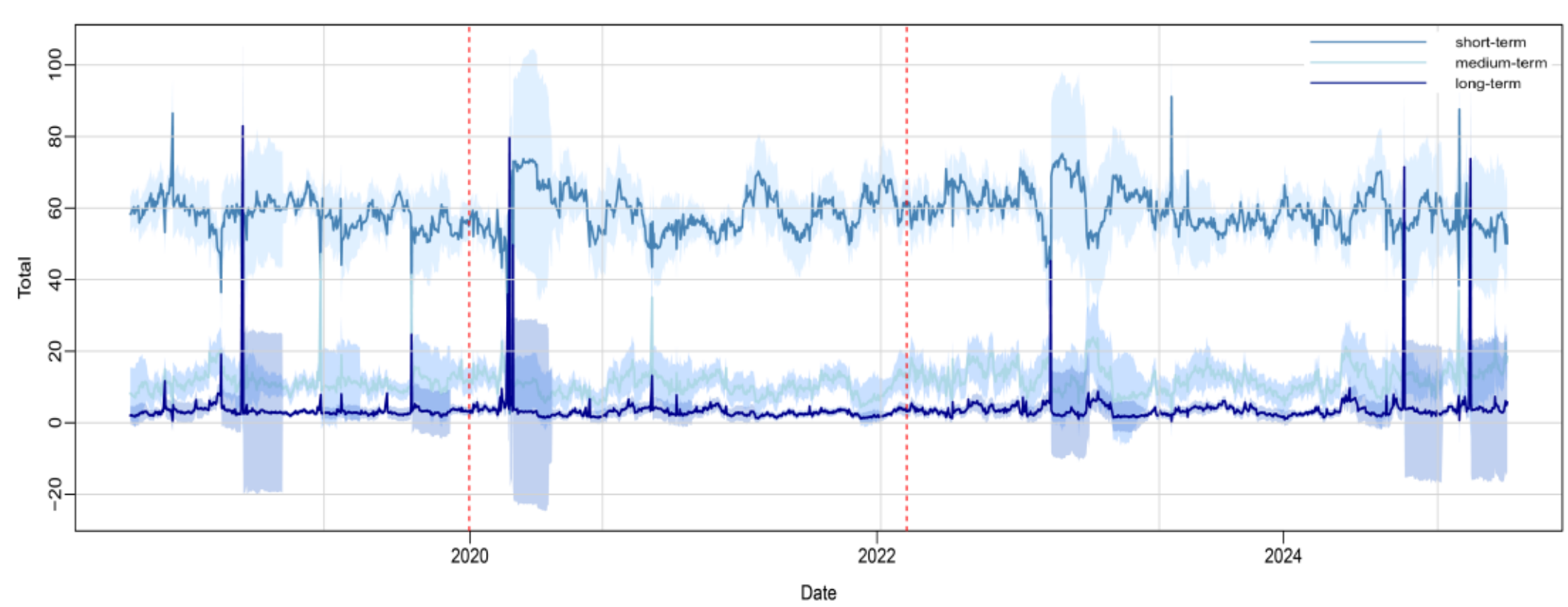

(b) Dynamic total connectedness on frequency decomposition

Figure 3. Dynamic total connectedness of return series. The connectedness is estimated using the Bootstrap method with a 95% confidence interval, and the shaded areas represent the confidence bounds.

Figure 3(b) shows dynamic total connectedness for short-term (within 5 days), medium-term (6th to 22nd days) and long-term (23rd days and later). As for the short-term connectedness, financial markets seem to process information quickly and market shocks mainly affect short-term cyclical behaviour. In contrast, the variation of long-term connectedness may be due to a fundamental change in investor expectations which has a long-term impact on systemic risk. These expectations are then transmitted to the related assets in the portfolio. According to Figure 3(b), the short-term connectedness dominates both the medium-term and long-term connectedness for the whole period.

The reason is likely to be that markets were calmer with relatively low uncertainty prior to the market shock, and investors were less fearful of market movements. The shocks that create future uncertainty in the system propagate more quickly and their impact on the system diminishes after a few days, resulting in short-term connectedness. However, a significant increase in long-term connectedness occurs in late 2022 and late 2024. The possible reason is the high level of volatility in the financial system in general due to the continued impact and uncertainty of the pandemic, as well as the economic situation. In this context, stock market prices have been falling and volatility has been increasing. The increasing uncertainty leads to more persistent responses to shocks by investors. The result of high connectedness driven by low-frequency responses to shocks leads to long-term uncertainty, causing systemic risk to grow during these periods.

Table 4 presents the GFEVD-based average joint connectedness results for the return series. The connectedness based on frequency decomposition is shown in Appendix. Table 4 contains the contributions to one's own return of the cryptocurrencies and the green financial indices in the diagonal. According to the results, after the occurrence of the COVID-19, the intrinsic contribution of GBI to its own return decreases from approximately 89.72% to 49.95%. And the value increases a little to 64.16% after the outbreak of Russia-Ukraine war. This shift demonstrates that each asset obtains greater contributions from other assets after the market shocks. This outcome indicates that the green financial market is readily impacted by external shocks.

Moreover, considering the influences of one asset to (from) others, it is noticeable that The Total Connectedness Index (TCI) significantly increases from 61.92% in the Pre COVID-19 period to 69.87% during the COVID-19 pandemic, reflecting heightened market interconnectedness and intensified risk transmission between cryptocurrencies and green financial markets amid global uncertainty. However, during the Russia-Ukraine war period, the TCI notably drops to 54.11%, indicating weakened interdependencies and potentially signalling investors' different risk perceptions and increased hedging behaviors amid geopolitical tensions. Analyzing specific cryptocurrencies, traditional cryptocurrency (BTC) exhibits increased vulnerability during COVID-19, with its FROM connectedness rising from 71.78% to 76.28%, signifying greater exposure to systemic risks. However, during the Russia-Ukraine war, BTC’s FROM value decreases substantially to 65.12%, suggesting that it provided improved hedging capabilities in geopolitical crisis conditions. For sustainable cryptocurrencies (XRP, ADA, XLM, MIOTA, POWR, SNC), most show heightened connectedness and greater susceptibility to market risks during COVID-19, except for XRP which slightly decreases. Conversely, during the Russia-Ukraine war, all sustainable cryptocurrencies demonstrate significantly lower risk transmission, particularly MIOTA and POWR, thus revealing their potential utility as hedging assets and portfolio diversifiers in geopolitical crises. Therefore, the Russia-Ukraine war provides clearer evidence of sustainable cryptocurrencies' and traditional cryptocurrency's potential roles as effective hedging instruments. Investors aiming for diversified portfolios could particularly consider MIOTA and POWR due to their lower connectedness and stronger diversification capabilities during geopolitical instability.

Table 4: Joint connectedness of return series

| | BTC | XRP | ADA | XLM | MIOTA | POWR | SNC | GBI | ESG | CEI | WIND | SOLAR | **FROM** |
|---|---|---|---|---|---|---|---|---|---|---|---|---|---|
| **Pre_COVID-19** | | | | | | | | | | | | | |
| BTC | 28.22 | 13.03 | 14.89 | 12.53 | 12.75 | 11.37 | 7.16 | 0.01 | 0.00 | 0.00 | 0.01 | 0.02 | 71.78 |
| XRP | 12.10 | 26.31 | 16.25 | 14.84 | 14.19 | 11.21 | 4.31 | 0.01 | 0.25 | 0.25 | 0.07 | 0.19 | 73.69 |
| ADA | 12.47 | 14.72 | 23.77 | 15.84 | 14.40 | 13.14 | 5.13 | 0.00 | 0.12 | 0.12 | 0.12 | 0.17 | 76.23 |
| XLM | 11.41 | 14.49 | 17.04 | 25.63 | 12.71 | 12.52 | 5.53 | 0.07 | 0.16 | 0.16 | 0.12 | 0.17 | 74.37 |
| MIOTA | 11.73 | 14.05 | 15.75 | 12.89 | 25.99 | 12.45 | 6.90 | 0.01 | 0.06 | 0.07 | 0.06 | 0.05 | 74.01 |
| POWR | 10.91 | 11.54 | 14.97 | 13.20 | 12.96 | 27.05 | 8.51 | 0.07 | 0.23 | 0.24 | 0.17 | 0.13 | 72.95 |
| SNC | 10.31 | 6.44 | 8.75 | 8.78 | 10.82 | 12.99 | 41.12 | 0.03 | 0.15 | 0.18 | 0.32 | 0.12 | 58.88 |
| GBI | 0.09 | 0.44 | 0.47 | 0.14 | 0.07 | 0.23 | 0.06 | 89.72 | 0.85 | 0.88 | 6.01 | 1.04 | 10.28 |
| ESG | 0.36 | 0.71 | 0.55 | 0.81 | 0.15 | 0.59 | 0.25 | 0.56 | 36.26 | 36.19 | 12.36 | 11.21 | 63.74 |
| CEI | 0.35 | 0.69 | 0.53 | 0.78 | 0.15 | 0.58 | 0.26 | 0.54 | 36.05 | 36.26 | 12.55 | 11.26 | 63.74 |
| WIND | 0.62 | 1.26 | 1.11 | 0.96 | 0.39 | 0.81 | 0.60 | 3.40 | 17.92 | 18.35 | 44.31 | 10.26 | 55.69 |
| SOLAR | 0.47 | 1.07 | 0.69 | 0.72 | 0.10 | 0.44 | 0.17 | 0.52 | 16.44 | 16.62 | 10.41 | 52.37 | 47.63 |
| **TO** | 70.82 | 78.44 | 91.00 | 81.49 | 78.69 | 76.33 | 38.88 | 5.22 | 72.23 | 73.06 | 42.20 | 34.62 | **TCI=** |
| **NET** | -0.96 | 4.75 | 14.77 | 7.12 | 4.68 | 3.38 | -20.00 | -5.06 | 8.49 | 9.32 | -13.49 | -13.01 | **61.92** |
| **During_COVID-19** | | | | | | | | | | | | | |
| BTC | 23.72 | 7.78 | 12.44 | 10.72 | 12.13 | 8.29 | 8.21 | 1.10 | 4.39 | 4.44 | 3.45 | 3.34 | 76.28 |
| XRP | 9.63 | 29.24 | 12.03 | 17.31 | 12.82 | 6.76 | 4.63 | 0.41 | 2.32 | 2.34 | 1.24 | 1.27 | 70.76 |
| ADA | 12.83 | 10.01 | 24.27 | 13.99 | 13.85 | 7.31 | 5.23 | 0.76 | 3.63 | 3.66 | 2.30 | 2.15 | 75.73 |
| XLM | 11.22 | 14.54 | 14.06 | 24.53 | 12.93 | 7.35 | 5.60 | 0.59 | 2.74 | 2.77 | 2.03 | 1.65 | 75.47 |
| MIOTA | 12.27 | 10.42 | 13.53 | 12.60 | 24.08 | 10.08 | 5.77 | 0.41 | 3.30 | 3.31 | 2.18 | 2.05 | 75.92 |
| POWR | 11.17 | 7.28 | 9.28 | 9.41 | 13.51 | 32.58 | 6.08 | 0.27 | 3.15 | 3.16 | 2.21 | 1.91 | 67.42 |
| SNC | 13.04 | 6.30 | 8.18 | 8.72 | 9.14 | 7.18 | 38.29 | 0.48 | 2.67 | 2.72 | 1.87 | 1.41 | 61.71 |
| GBI | 2.91 | 0.81 | 2.14 | 1.38 | 1.17 | 0.53 | 0.84 | 49.95 | 10.42 | 10.66 | 12.36 | 6.83 | 50.05 |
| ESG | 4.78 | 1.99 | 3.83 | 2.84 | 3.56 | 2.56 | 1.82 | 3.40 | 25.80 | 25.72 | 13.39 | 10.31 | 74.20 |
| CEI | 4.79 | 1.99 | 3.83 | 2.85 | 3.55 | 2.57 | 1.84 | 3.45 | 25.51 | 25.57 | 13.58 | 10.48 | 74.43 |
| WIND | 4.18 | 1.18 | 2.82 | 2.45 | 2.57 | 1.96 | 1.55 | 5.87 | 15.54 | 15.91 | 29.47 | 16.50 | 70.53 |
| SOLAR | 4.74 | 1.46 | 3.03 | 2.42 | 2.81 | 1.98 | 1.52 | 2.66 | 13.44 | 13.79 | 18.11 | 34.04 | 65.96 |
| **TO** | 91.56 | 63.76 | 85.17 | 84.69 | 88.04 | 56.57 | 43.09 | 19.40 | 87.11 | 88.48 | 72.72 | 57.90 | **TCI=** |
| **NET** | 15.28 | -7.00 | 9.44 | 9.22 | 12.12 | -10.85 | -18.62 | -30.65 | 12.91 | 14.05 | 2.19 | -8.06 | **69.87** |
| **During_R-U war** | | | | | | | | | | | | | |
| BTC | 34.88 | 6.71 | 15.17 | 6.21 | 10.51 | 6.58 | 16.42 | 0.15 | 2.43 | 0.40 | 0.32 | 0.22 | 65.12 |
| XRP | 6.96 | 36.05 | 14.35 | 19.95 | 10.56 | 3.83 | 4.31 | 0.10 | 1.94 | 0.86 | 0.56 | 0.54 | 63.95 |
| ADA | 13.23 | 11.91 | 30.24 | 11.24 | 13.70 | 7.33 | 8.33 | 0.05 | 2.36 | 0.42 | 0.67 | 0.52 | 69.76 |
| XLM | 6.40 | 19.92 | 14.44 | 35.91 | 11.08 | 4.57 | 4.01 | 0.20 | 1.52 | 0.70 | 0.72 | 0.54 | 64.09 |
| MIOTA | 10.07 | 10.44 | 15.94 | 10.24 | 33.97 | 6.83 | 8.75 | 0.21 | 1.59 | 0.92 | 0.47 | 0.57 | 66.03 |
| POWR | 9.09 | 5.18 | 12.50 | 6.14 | 9.86 | 47.80 | 4.45 | 0.53 | 2.26 | 1.19 | 0.82 | 0.19 | 52.20 |
| SNC | 20.28 | 4.99 | 11.51 | 4.71 | 10.76 | 3.64 | 42.21 | 0.01 | 1.44 | 0.01 | 0.35 | 0.07 | 57.79 |
| GBI | 0.28 | 0.09 | 0.52 | 0.23 | 0.45 | 0.61 | 0.03 | 64.16 | 5.39 | 0.43 | 19.57 | 8.25 | 35.84 |
| ESG | 3.19 | 2.68 | 3.95 | 2.16 | 2.50 | 2.13 | 1.65 | 3.75 | 49.40 | 1.31 | 15.78 | 11.50 | 50.60 |
| CEI | 0.85 | 1.85 | 0.96 | 1.40 | 2.28 | 2.02 | 0.16 | 0.74 | 1.75 | 81.49 | 2.21 | 4.31 | 18.51 |
| WIND | 0.42 | 0.52 | 1.08 | 0.70 | 0.69 | 0.59 | 0.34 | 13.90 | 14.78 | 1.17 | 44.83 | 20.98 | 55.17 |
| SOLAR | 0.22 | 0.66 | 0.73 | 0.76 | 0.81 | 0.23 | 0.03 | 7.15 | 12.83 | 2.54 | 24.28 | 49.75 | 50.25 |
| **TO** | 70.99 | 64.95 | 91.15 | 63.74 | 73.20 | 38.36 | 48.48 | 26.79 | 48.29 | 9.95 | 65.75 | 47.69 | **TCI=** |
| **NET** | 5.87 | 1.00 | 21.39 | -0.35 | 7.17 | -13.84 | -9.31 | -9.05 | -2.31 | -8.56 | 10.58 | -2.56 | **54.11** |

Notes: The results depend on a 20-step-ahead GFEVD and a 2 lag TVP-VAR model (chosen by the minimum AIC value).

Considering the pairwise connectedness of return series, which are the off-diagonal elements in Table 4 the results indicate that the information spillovers between cryptocurrencies and green financial market are relatively low for the whole period. The fact that movements in one market are less susceptible to movements in another lays the foundation for diversification benefits of our portfolios containing sustainable cryptocurrencies. The same result can be found in Li and Meng (2022). Particularly, in the face of market shocks, such as the COVID-19, the connectedness between the cryptocurrency market and the green financial market has increased, but remains at a relatively low level. Therefore, the inclusion of cryptocurrencies in a portfolio objectively serves as an effective hedge against market shocks, mitigating risk and enhancing overall portfolio performance. Another interesting result is that the connectedness of sustainable cryptocurrencies to stocks is lower than the connectedness of Bitcoin to stocks. For example, during the Russia-Ukraine war period, the pairwise connectedness between BTC and GBI is 0.28% while the pairwise connectedness between XRP and GBI is 0.09%. Therefore, in the following parts we will further explore whether sustainable cryptocurrencies are more suitable hedging tools in portfolios than conventional cryptocurrencies.

Ultimately, for a comprehensive evaluation of information spillover between cryptocurrency and green financial market, Figure 4 illustrates the connectedness network of their return series. It can be found that spillovers among assets within the same market, particularly among green financial assets, are stronger. Shown in the Figure 4, GBI is the main source of information spillover to ESG and WIND. It is worth noting that the lower degree of connectedness between cryptocurrencies and green financial indices may provide diversification benefits in investment portfolio. The direction of spillovers can be further explored by observing the direction of the arrows in the figure. In detail, BTC, XLM, ADA, MIOTA, ESG and WIND are net receivers while XRP, POWR, SNC, CEI, GBI and SOLAR are net transmitters in the connectedness network. This is consistent with the findings of Lu et al. (2023), who identified CEI as a major transmitter in the volatility-connectedness network, highlighting its critical role in facilitating financial contagion. In a similar vein, Chen et al. (2022a) argued that clean energy assets have emerged as systemic sources of spillover transmission since 2015, largely due to momentum gained after the Paris Agreement.

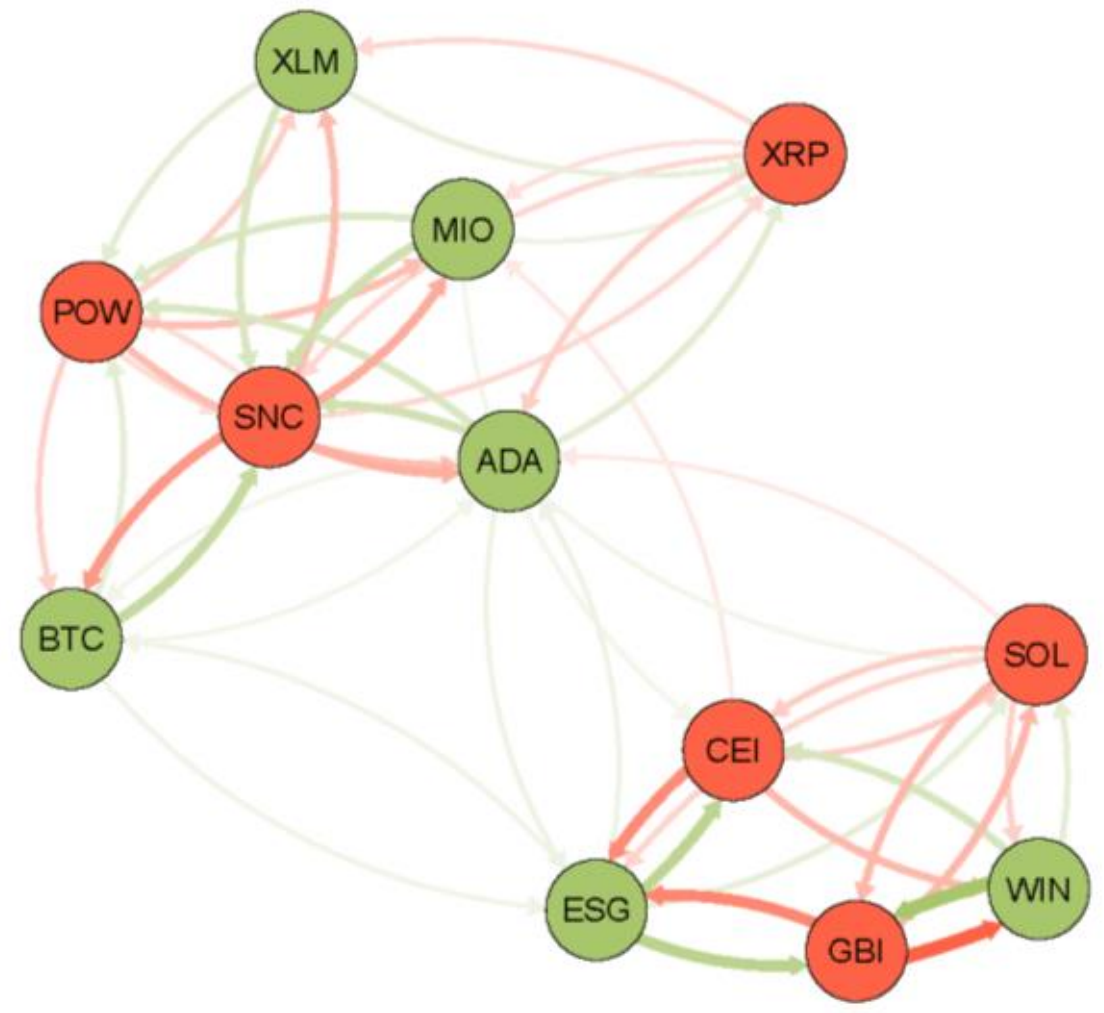

Figure 4. The network of return series. The red (green) node indicates the net transmitter (receiver) of the spillover system. The strength of the net pairwise spillover is indicated by the thickness of the edge arrow.

## *5.2 Connectedness analysis of volatility series*

Markets with high volatility connectedness mean the risks that each market suffer is correlated. Volatility connectedness affords investors the capability to promptly adjust their asset allocation choices. Therefore, we adopt connectedness analysis for the volatility series.

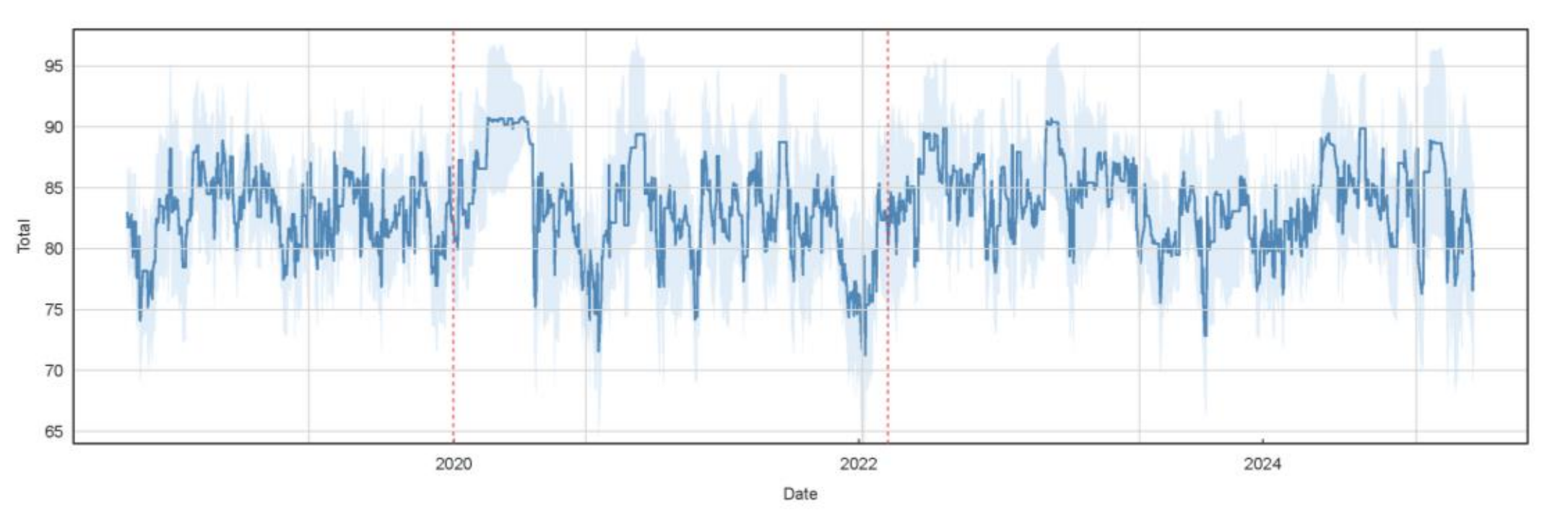

Figure 5. Dynamic total connectedness of volatility series. The connectedness is estimated using the Bootstrap method with a 95% confidence interval, and the shaded areas represent the confidence bounds.

Figure 5 mainly shows a visual depiction of the dynamic connectedness over the whole period. It can be seen that the overall connectedness are significantly higher, ranging from 72% to 91%. Unlike the return connectedness, the total connectedness of the volatility series exhibits greater fluctuation throughout the study period. The results indicate that when extreme events are encountered, there is a sudden increase in the degree of integration between the volatility series. For example, in 2020, the total volatility spillover starts to pick up. The reason is likely to be the turbulence in the cryptocurrency market. For example, in 2022, the cryptocurrency market shows a clear downward trend. Many cryptocurrencies have fallen back significantly from their all-time highs, with the value of major cryptocurrencies, including Bitcoin and Ethereum plummeting over the course of the year.

The GFEVD-based average joint connectedness results for the volatility series are given in

Table 5. The own volatility contributions of the cryptocurrencies and the green financial indices are shown in the diagonal in

Table 5. From the results, the self-contribution of cryptocurrencies always keeps a relatively high level both before and during the market shock periods, indicating their independence of volatility. Regarding the volatility spillover among the assets, during the COVID-19 period, three assets BTC, XLM and WIND are the main contributors to the whole system, with 78.29%, 81.66% and 86.42%, respectively. However, during the Russia-Ukraine war period, the main contributors to the whole system have changed to XRP, XLM and MIOTA, with 79.74%, 79.75% and 80.49%, respectively. It has shown that the role of each asset for volatility spillover changes a lot.

Table 5: Joint connectedness of volatility series

| | BTC | XRP | ADA | XLM | MIOTA | POWR | SNC | GBI | ESG | CEI | WIND | SOLAR | **FROM** |
|---|---|---|---|---|---|---|---|---|---|---|---|---|---|
| **Pre_COVID-19** | | | | | | | | | | | | | |
| BTC | 55.24 | 2.10 | 6.79 | 3.94 | 8.06 | 4.03 | 5.51 | 0.71 | 0.98 | 0.81 | 8.84 | 2.99 | 44.76 |
| XRP | 3.72 | 30.93 | 8.98 | 4.93 | 22.75 | 1.41 | 16.43 | 2.20 | 0.50 | 0.67 | 2.80 | 4.69 | 69.07 |
| ADA | 10.99 | 5.37 | 22.93 | 9.97 | 23.11 | 5.86 | 6.45 | 1.01 | 5.28 | 3.70 | 2.66 | 2.66 | 77.07 |
| XLM | 4.86 | 4.99 | 9.75 | 38.73 | 11.76 | 5.57 | 1.17 | 6.26 | 4.85 | 3.52 | 3.56 | 4.98 | 61.27 |
| MIOTA | 7.75 | 3.48 | 7.10 | 10.58 | 40.33 | 1.29 | 11.26 | 1.14 | 4.30 | 3.05 | 7.90 | 1.81 | 59.67 |
| POWR | 3.70 | 1.42 | 11.20 | 7.89 | 9.61 | 29.47 | 0.83 | 3.90 | 5.78 | 4.53 | 15.50 | 6.17 | 70.53 |
| SNC | 3.96 | 5.67 | 1.90 | 0.57 | 1.47 | 5.09 | 72.80 | 3.17 | 1.48 | 1.18 | 1.05 | 1.64 | 27.20 |
| GBI | 9.07 | 0.16 | 5.14 | 2.10 | 2.31 | 3.81 | 1.11 | 67.44 | 0.39 | 0.28 | 5.27 | 2.93 | 32.56 |
| ESG | 3.00 | 9.81 | 0.10 | 1.79 | 3.92 | 1.27 | 7.26 | 5.42 | 26.86 | 25.36 | 2.59 | 12.62 | 73.14 |
| CEI | 2.98 | 9.59 | 0.13 | 1.73 | 3.70 | 1.36 | 7.18 | 6.04 | 26.22 | 24.99 | 2.91 | 13.18 | 75.01 |
| WIND | 1.13 | 6.78 | 1.76 | 5.91 | 10.82 | 0.84 | 6.81 | 7.46 | 8.28 | 7.47 | 32.35 | 10.39 | 67.65 |
| SOLAR | 0.78 | 4.69 | 0.42 | 1.74 | 4.13 | 3.31 | 5.72 | 16.81 | 10.96 | 10.38 | 7.80 | 33.27 | 66.73 |
| **TO** | 51.94 | 54.05 | 53.27 | 51.15 | 101.64 | 33.83 | 69.72 | 54.12 | 69.03 | 60.96 | 60.88 | 64.06 | **TCI=** |
| **NET** | 7.18 | -15.02 | -23.80 | -10.12 | 41.97 | -36.70 | 42.52 | 21.56 | -4.11 | -14.05 | -6.77 | -2.67 | **60.39** |
| **During_COVID-19** | | | | | | | | | | | | | |
| BTC | 21.71 | 8.20 | 11.88 | 8.81 | 8.13 | 9.20 | 1.04 | 6.41 | 7.61 | 7.39 | 3.59 | 6.04 | 78.29 |
| XRP | 8.59 | 47.67 | 5.90 | 6.45 | 6.04 | 3.54 | 0.47 | 14.60 | 1.42 | 1.41 | 0.26 | 3.65 | 52.33 |
| ADA | 11.68 | 10.52 | 25.12 | 11.28 | 6.31 | 6.50 | 0.34 | 15.68 | 4.49 | 4.14 | 1.48 | 2.47 | 74.88 |
| XLM | 9.66 | 31.49 | 7.60 | 18.34 | 4.97 | 2.78 | 0.88 | 17.30 | 1.85 | 1.72 | 0.71 | 2.70 | 81.66 |
| MIOTA | 15.41 | 11.17 | 16.27 | 6.10 | 22.29 | 9.36 | 0.70 | 7.55 | 3.91 | 3.72 | 1.42 | 2.10 | 77.71 |
| POWR | 11.10 | 3.09 | 4.93 | 4.08 | 10.78 | 44.81 | 4.69 | 3.49 | 2.86 | 2.83 | 3.97 | 3.37 | 55.19 |
| SNC | 2.72 | 1.78 | 4.99 | 4.58 | 3.41 | 11.42 | 25.93 | 8.56 | 15.70 | 15.63 | 2.07 | 3.22 | 74.07 |
| GBI | 14.36 | 3.50 | 1.35 | 0.24 | 1.85 | 3.72 | 4.92 | 30.15 | 10.36 | 10.23 | 6.07 | 13.27 | 69.85 |
| ESG | 13.08 | 0.99 | 3.67 | 0.82 | 0.38 | 10.62 | 4.18 | 1.24 | 24.91 | 24.30 | 6.70 | 9.11 | 75.09 |
| CEI | 13.39 | 0.85 | 3.92 | 0.92 | 0.44 | 10.89 | 3.98 | 1.30 | 24.57 | 24.03 | 6.73 | 8.99 | 75.97 |
| WIND | 13.24 | 0.36 | 4.66 | 2.70 | 0.96 | 8.40 | 2.96 | 1.99 | 19.51 | 19.28 | 13.58 | 12.36 | 86.42 |
| SOLAR | 8.77 | 0.09 | 4.44 | 1.50 | 0.35 | 4.95 | 3.74 | 2.85 | 20.13 | 19.67 | 10.26 | 23.26 | 76.74 |
| **TO** | 122.00 | 72.04 | 69.61 | 47.48 | 43.62 | 81.38 | 27.90 | 80.97 | 112.41 | 110.32 | 43.26 | 67.28 | **TCI=** |
| **NET** | 43.71 | 19.71 | -5.27 | -34.18 | -34.09 | 26.19 | -46.17 | 11.12 | 37.32 | 34.35 | -43.16 | -9.46 | **73.18** |
| **During_R-U war** | | | | | | | | | | | | | |
| BTC | 25.34 | 0.50 | 4.45 | 3.39 | 7.05 | 15.56 | 16.97 | 3.26 | 5.10 | 5.21 | 0.35 | 12.84 | 74.66 |
| XRP | 5.07 | 20.26 | 10.70 | 14.43 | 12.27 | 21.65 | 2.30 | 4.25 | 1.32 | 0.47 | 0.83 | 6.45 | 79.74 |
| ADA | 5.62 | 5.07 | 23.55 | 16.13 | 4.46 | 19.36 | 5.11 | 0.25 | 1.32 | 0.93 | 1.10 | 17.11 | 76.45 |
| XLM | 2.88 | 6.92 | 12.83 | 20.25 | 12.52 | 29.95 | 2.46 | 1.56 | 2.43 | 0.78 | 0.83 | 6.58 | 79.75 |
| MIOTA | 0.72 | 0.78 | 10.84 | 16.37 | 19.51 | 23.77 | 7.02 | 1.26 | 1.47 | 0.96 | 1.14 | 16.17 | 80.49 |
| POWR | 0.43 | 0.30 | 15.67 | 17.86 | 14.62 | 37.41 | 5.48 | 2.18 | 2.51 | 0.40 | 1.69 | 1.45 | 62.59 |
| SNC | 7.79 | 1.70 | 3.28 | 1.61 | 13.33 | 7.09 | 47.34 | 6.88 | 3.10 | 0.36 | 0.30 | 7.22 | 52.66 |
| GBI | 0.49 | 0.44 | 2.26 | 3.40 | 6.86 | 4.82 | 2.44 | 35.64 | 10.83 | 5.13 | 21.96 | 5.74 | 64.36 |
| ESG | 0.33 | 1.01 | 17.97 | 14.51 | 7.13 | 31.87 | 0.06 | 2.40 | 21.78 | 0.21 | 1.41 | 1.31 | 78.22 |
| CEI | 3.34 | 0.47 | 5.62 | 7.92 | 10.64 | 12.95 | 9.99 | 2.22 | 4.88 | 33.70 | 4.71 | 3.56 | 66.30 |
| WIND | 1.18 | 2.51 | 1.50 | 2.41 | 10.51 | 11.79 | 0.46 | 7.55 | 15.90 | 1.52 | 36.89 | 7.78 | 63.11 |
| SOLAR | 2.43 | 2.26 | 0.79 | 5.05 | 11.98 | 12.36 | 0.89 | 4.57 | 5.37 | 1.28 | 16.56 | 36.46 | 63.54 |
| **TO** | 30.28 | 21.96 | 85.91 | 103.08 | 111.37 | 191.17 | 53.18 | 36.38 | 54.23 | 17.25 | 50.88 | 86.21 | **TCI=** |
| **NET** | -44.38 | -57.78 | 9.46 | 23.33 | 30.88 | 128.58 | 0.52 | -27.98 | -23.99 | -49.05 | -12.23 | 22.67 | **70.16** |

Notes: The results depend on a 20-step-ahead GFEVD and a 1 lag TVP-VAR model (chosen by the minimum AIC value).

Considering the pairwise connectedness of volatility series, which are the off-diagonal elements in Table 5, the results are consistent with the observation of return series spillover (shown in Table 4).

It shows that although the pairwise connectedness between cryptocurrencies and green financial indices increases during the COVID-19 pandemic, it remains at a relatively low level. However, the volatility spillovers between BTC and six types of sustainable cryptocurrencies decrease after the Russia-Ukraine war outbreak, indicating that the sustainable cryptocurrencies and the traditional cryptocurrencies do not share the same impact of risks when experiencing global energy crises.

### *5.3 Robustness check*

In this section, we verify the robustness of our experiments by varying the parameters of the TVP-VAR model and comparing them with the results of the Markov transformation model.

#### *5.3.1 Sensitivity analysis: parameter variation*

The robustness check in this part is to verify the primary results by changing forecast step. Specifically, in addition to the twenty steps previously employed, we choose two and ten steps as forecasting steps to measure the dynamic joint total connectedness outcomes of volatility and returns, respectively. If adjustments to the forecast step have no discernible impact on our findings, then the main conclusions in the study are verified to be robust. Figure 6 displays the results of the robustness verification. It can be seen that the dynamic joint total connectedness decreases with decreasing forecast steps, with a range of 2, 10, and 20. However, the total trends of these three approaches are the same. In particular, early 2020 shows a notable increase in total connectedness across all robustness tests, confirming once more the effect of the COVID-19 pandemic on total connectedness. With respect to the return series, the results do not show significant differences from the three distinct forecast steps, suggesting that findings in this study are reliable. Furthermore, the three lines in the volatility series exhibit greater variation than those in the return series, but their differences still remain within approximately 15%. Therefore, it can be inferred that the dynamic joint total connectedness is not greatly affected by the forecast step, and the findings are independent of the step length.

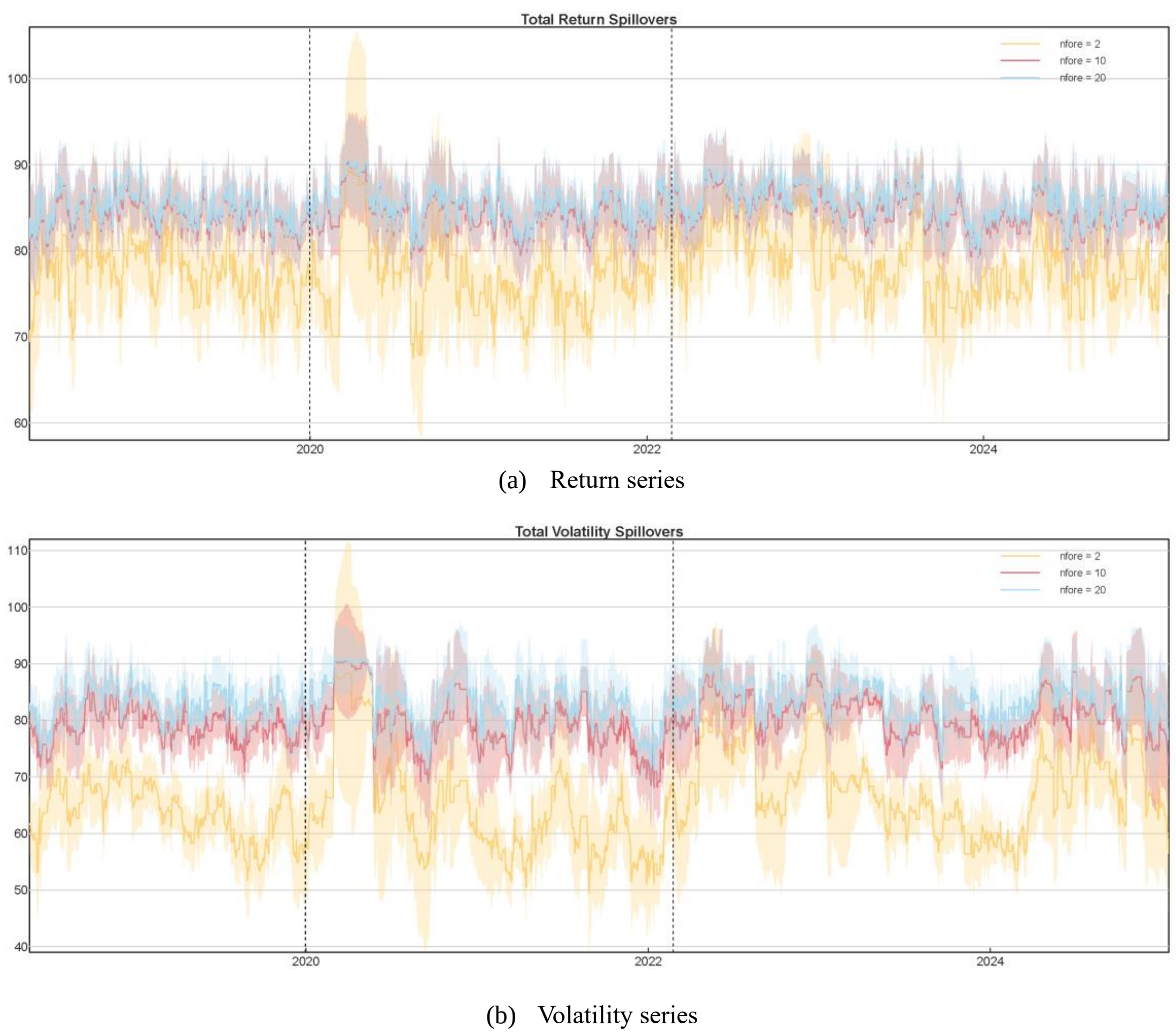

(a) Return series

(b) Volatility series

Figure 6. Robustness check for return and volatility series. The connectedness is estimated using the Bootstrap method with a 95% confidence interval, and the shaded areas represent the confidence bounds.

### *5.3.2 Comparison with Markov Switching Model*

The Markov Switching Model (MSM) is a statistical model that particularly useful for capturing structural shifts in time series data, such as economic cycles, financial market fluctuations, or other dynamic processes that exhibit regime-dependent behaviors (Kim and Nelson, 1999). In our study, we compare the results of the TVP-VAR model with those obtained from the MSM to assess the robustness of our findings. The TVP-VAR model accounts for gradual changes in parameters over time, whereas the MSM explicitly models abrupt regime shifts.

Table 6: Markov Switching Model Estimation Results

| | | Regime 1 | Regime 2 |
|---|---|---|---|
| BTC | ESG | 0.58*** | 2.60*** |
| | CEI | 0.06** | -1.42** |
| | SOLAR | 0.12* | 0.16 |
| XRP | ESG | 0.78*** | 2.72* |
| | CEI | 0.17*** | 0.14 |
| | WIND | -0.25* | -0.23 |
| ADA | ESG | 3.22*** | 1.01*** |
| | CEI | -1.06 | 0.08* |
| XLM | ESG | 0.78*** | 1.61 |
| | CEI | 0.12** | 0.60 |
| MIOTA | GBI | 0.20*** | -0.50 |
| | ESG | 1.01*** | 2.44* |
| | CEI | 0.18*** | -0.40 |
| POWR | ESG | 0.79*** | 3.33* |
| | CEI | 0.22*** | -0.89 |
| SNC | ESG | 1.11 | 1.09*** |

Notes: The table presents only statistically significant results. Regime 1 represents the stable market period, while Regime 2 represents the turbulent market period.

As shown in Table 6, Regime 1 represents the stable market period, while Regime 2 represents the turbulent market period. The table only presents significant results. During the stable market period (Regime 1), the impact of the green financial market is generally low, which aligns with the TVP-VAR results, indicating that green financial factors have a relatively weak association with cryptocurrencies most of the time. However, during the turbulent market period (Regime 2), the influence of certain green financial variables (such as ESG and CEI) on cryptocurrencies increases. Specifically, the impact of ESG on BTC rises from 0.58 in Regime 1 to 2.60 in Regime 2. The influence of ESG on cryptocurrencies such as XRP, ADA, and POWR also significantly strengthens in Regime 2. This finding is consistent with the TVP-VAR results, which suggest that the pair connectedness between cryptocurrencies and the green financial market intensified during the COVID-19 period. This robustness check enhances the credibility of our conclusions and provides a more comprehensive understanding of the underlying dynamics in our data.

## 6. Hedging performance analysis

In this section, the DCC and Copula models provided the estimation results which can be employed to design the optimal portfolio. This analysis seems to be relevant and decisive for international investors and for risk management triggered by price volatility of cryptocurrencies and green financial indices. Our strategy entails constructing a hedged portfolio comprising diverse variables. Specifically, we assume that an investor allocates investments in cryptocurrencies as a means of hedging against potential turbulence emanating from the green financial markets. Thus, we assume that the investor's goal is to minimize the risk of a portfolio consisting of seven types of cryptocurrencies and five green financial indices.

Figure 7 shows the estimated results of optimal hedging ratios by DCC model and Copula model. The dynamic plot of hedging ratio shows that the two multivariate GARCH models exhibit some consistency in the estimation of optimal hedging ratio. The optimal hedging ratio is maintained at a steady state until the global pandemic emerges. However, with the rapid spread of COVID-19, the optimal hedging ratio reaches its peak in the early 2020s. And there are sharp fluctuations during the COVID-19 period. Meanwhile, the estimation results of both DCC-GARCH and Copula-GARCH models show that the optimal hedging ratio tends to increase with the increasing market uncertainty caused by Russia-Ukraine war. By comparing the seven cryptocurrencies, the optimal hedging ratio of BTC changes more than that of other six sustainable cryptocurrencies, indicating that it is more affected by the uncertainty caused by market shocks. Among the portfolios of cryptocurrencies with the five green financial indices, the one containing GBI has the smallest optimal hedging ratio and the lowest volatility.

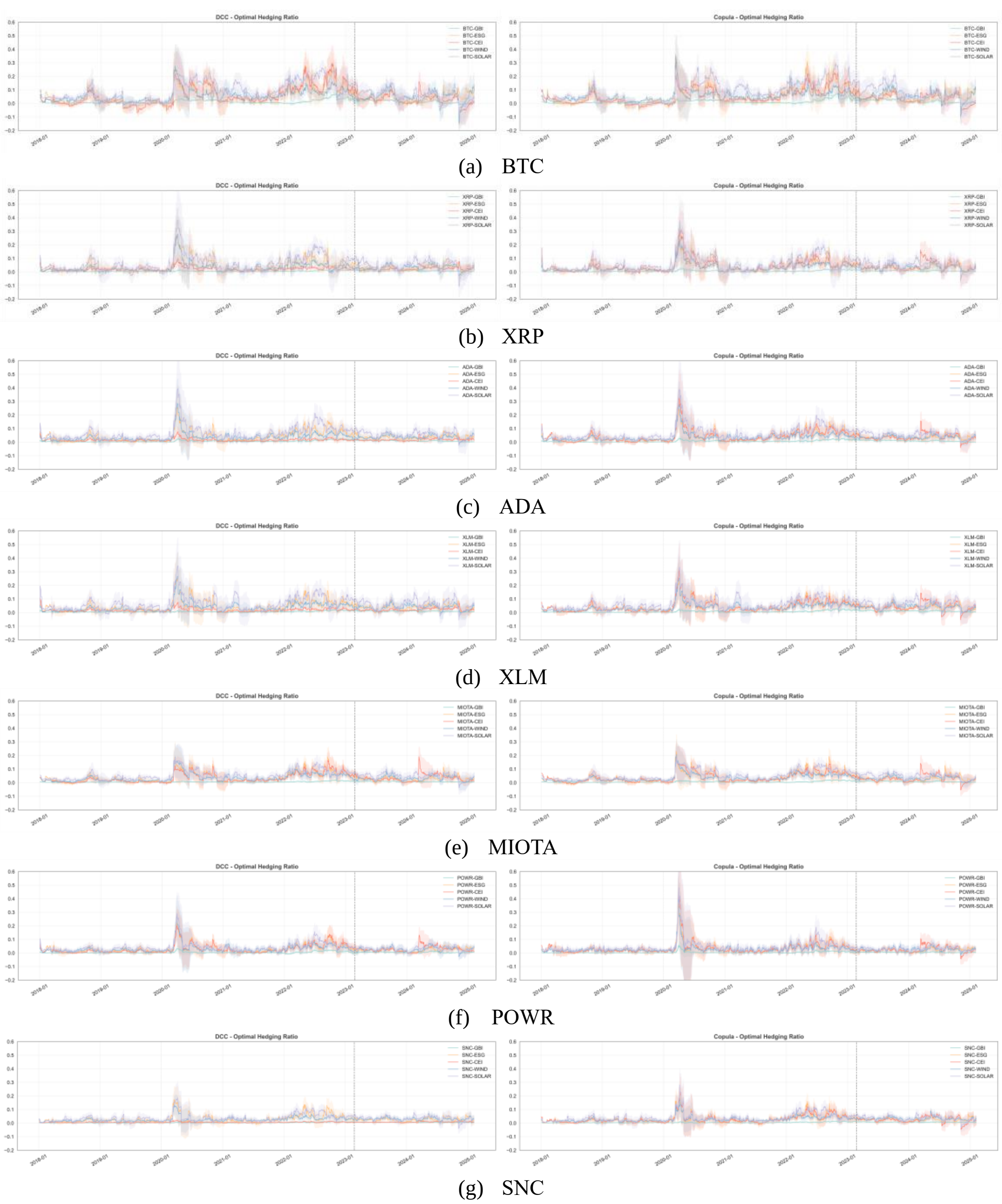

(a) BTC

(b) XRP

(c) ADA

(d) XLM

(e) MIOTA

(f) POWR

(g) SNC

Figure 7. Optimal hedging ratio estimated by DCC-GARCH model (left) and Copula-GARCH model (right). The hedging ratio is estimated using the Bootstrap method with a 95% confidence interval, and the shaded areas represent

the confidence bounds.

Table 7: Average optimal hedging ratio

| | | **Pre_COVID-19** | | **During_COVID-19** | | **During_R-U war** | |
|---|---|---|---|---|---|---|---|
| | | DCC | Copula | DCC | Copula | DCC | Copula |
| BTC | GBI | 0.0039 | 0.0039 | 0.0124 | 0.0138 | 0.0175 | 0.0171 |
| | ESG | 0.0186 | 0.0170 | 0.0449 | 0.0460 | 0.0837 | 0.0785 |
| | CEI | 0.0076 | 0.0142 | 0.0271 | 0.0356 | 0.0911 | 0.0772 |
| | WIND | 0.0279 | 0.0310 | 0.0481 | 0.0536 | 0.0751 | 0.0672 |
| | SOLAR | 0.0378 | 0.0340 | 0.0649 | 0.0804 | 0.1367 | 0.1303 |
| XRP | GBI | 0.0034 | 0.0030 | 0.0071 | 0.0072 | 0.0062 | 0.0077 |
| | ESG | 0.0194 | 0.0168 | 0.0311 | 0.0321 | 0.0568 | 0.0526 |
| | CEI | 0.0167 | 0.0187 | 0.0241 | 0.0392 | 0.0233 | 0.0559 |
| | WIND | 0.0164 | 0.0176 | 0.0310 | 0.0333 | 0.0435 | 0.0428 |
| | SOLAR | 0.0316 | 0.0320 | 0.0512 | 0.0554 | 0.0810 | 0.0798 |
| ADA | GBI | 0.0027 | 0.0020 | 0.0086 | 0.0087 | 0.0089 | 0.0092 |
| | ESG | 0.0189 | 0.0165 | 0.0354 | 0.0338 | 0.0551 | 0.0520 |
| | CEI | 0.0123 | 0.0182 | 0.0175 | 0.0360 | 0.0168 | 0.0554 |
| | WIND | 0.0209 | 0.0202 | 0.0364 | 0.0344 | 0.0469 | 0.0443 |
| | SOLAR | 0.0378 | 0.0365 | 0.0565 | 0.0548 | 0.0835 | 0.0811 |
| XLM | GBI | 0.0040 | 0.0027 | 0.0091 | 0.0103 | 0.0074 | 0.0093 |
| | ESG | 0.0227 | 0.0205 | 0.0367 | 0.0367 | 0.0567 | 0.0536 |
| | CEI | 0.0147 | 0.0208 | 0.0224 | 0.0412 | 0.0199 | 0.0565 |
| | WIND | 0.0196 | 0.0203 | 0.0383 | 0.0396 | 0.0496 | 0.0487 |
| | SOLAR | 0.0404 | 0.0403 | 0.0613 | 0.0629 | 0.0807 | 0.0775 |
| MIOTA | GBI | 0.0033 | 0.0025 | 0.0071 | 0.0073 | 0.0062 | 0.0072 |
| | ESG | 0.0158 | 0.0148 | 0.0281 | 0.0280 | 0.0529 | 0.0509 |
| | CEI | 0.0107 | 0.0173 | 0.0327 | 0.0307 | 0.0491 | 0.0518 |
| | WIND | 0.0171 | 0.0186 | 0.0312 | 0.0313 | 0.0479 | 0.0457 |
| | SOLAR | 0.0317 | 0.0298 | 0.0490 | 0.0491 | 0.0690 | 0.0721 |
| POWR | GBI | 0.0010 | 0.0016 | 0.0045 | 0.0045 | 0.0051 | 0.0054 |
| | ESG | 0.0166 | 0.0158 | 0.0229 | 0.0218 | 0.0398 | 0.0400 |
| | CEI | 0.0104 | 0.0172 | 0.0273 | 0.0242 | 0.0403 | 0.0418 |
| | WIND | 0.0174 | 0.0181 | 0.0259 | 0.0258 | 0.0365 | 0.0355 |
| | SOLAR | 0.0283 | 0.0280 | 0.0337 | 0.0322 | 0.0529 | 0.0578 |
| SNC | GBI | 0.0022 | 0.0026 | 0.0054 | 0.0068 | 0.0042 | 0.0052 |
| | ESG | 0.0139 | 0.0130 | 0.0238 | 0.0248 | 0.0363 | 0.0366 |
| | CEI | 0.0043 | 0.0137 | 0.0075 | 0.0246 | 0.0062 | 0.0379 |
| | WIND | 0.0168 | 0.0171 | 0.0302 | 0.0317 | 0.0315 | 0.0340 |
| | SOLAR | 0.0236 | 0.0242 | 0.0336 | 0.0352 | 0.0451 | 0.0488 |

Table 8: Average hedging effectiveness

| | | During_COVID-19 | | During_R-U war | |
|---|---|---|---|---|---|
| | | DCC | Copula | DCC | Copula |
| BTC | GBI | 0.98% | 0.97% | 1.83% | 1.78% |
| | ESG | 5.47% | 5.46% | 14.02% | 13.81% |
| | CEI | -0.08% | -0.04% | 14.09% | 15.01% |
| | WIND | 0.04% | 0.06% | 8.22% | 8.40% |
| | SOLAR | -1.61% | -1.99% | 10.18% | 9.52% |
| XRP | GBI | 1.40% | 1.48% | 2.26% | 2.19% |
| | ESG | 6.71% | 6.67% | 13.92% | 13.09% |
| | CEI | 0.55% | 0.54% | 13.26% | 12.95% |
| | WIND | 1.80% | 1.15% | 4.38% | 4.42% |
| | SOLAR | 1.45% | 1.32% | 4.97% | 4.85% |
| ADA | GBI | 1.39% | 1.51% | 2.10% | 1.88% |
| | ESG | 9.18% | 8.91% | 10.55% | 10.77% |
| | CEI | 0.64% | 0.62% | 6.04% | 6.11% |
| | WIND | 2.67% | 2.55% | 3.24% | 3.37% |
| | SOLAR | 1.84% | 1.70% | 5.75% | 5.62% |
| XLM | GBI | 1.25% | 1.44% | 1.73% | 1.35% |
| | ESG | 5.65% | 5.74% | 4.42% | 4.29% |
| | CEI | 0.83% | 0.81% | 6.38% | 6.36% |
| | WIND | 1.74% | 1.78% | 3.58% | 3.57% |
| | SOLAR | 1.20% | 1.16% | 3.63% | 3.54% |
| MIOTA | GBI | 1.28% | 1.30% | 1.21% | 1.01% |
| | ESG | 7.51% | 7.31% | 12.35% | 12.18% |
| | CEI | 0.33% | 0.32% | 12.81% | 12.43% |
| | WIND | 1.16% | 0.97% | 5.40% | 5.46% |
| | SOLAR | 1.04% | 0.91% | 5.43% | 5.44% |
| POWR | GBI | 0.91% | 1.07% | 1.04% | 1.13% |
| | ESG | 7.68% | 7.62% | 10.93% | 9.56% |
| | CEI | 4.69% | 4.04% | 10.58% | 10.52% |
| | WIND | 1.39% | 1.83% | 3.75% | 3.26% |
| | SOLAR | 0.20% | 0.39% | 3.67% | 3.49% |
| SNC | GBI | 1.20% | 1.20% | 1.05% | 1.10% |
| | ESG | 3.32% | 2.68% | 9.65% | 8.61% |
| | CEI | -0.15% | -0.16% | 5.48% | 5.05% |
| | WIND | 0.24% | 0.20% | 3.27% | 3.50% |
| | SOLAR | -0.48% | -0.46% | 2.83% | 1.93% |

However, it should be noted that the high volatility of cryptocurrency markets presents limitations to the effectiveness of the optimization methods applied here. While the models provide useful hedging ratios, the inherent fluctuations and unpredictability of cryptocurrencies could impact the reliability of these estimates over the long term. This makes it challenging to maintain a stable optimal hedging ratio, especially in the face of abrupt market shocks or prolonged periods of high volatility. Therefore, although the models suggest an increase in hedging ratios during times of uncertainty, their performance might be less effective during extreme market conditions.

Table 7 presents the average of the optimal hedging ratios before and after the occurrence of market shocks under two multivariate GARCH models. The estimation results of the two multivariate GARCH models are almost the same. The results show that the optimal hedging ratio of BTC/CEI is 0.0911 (calculated by DCC model) during the Russia-Ukraine war period, meaning that a $1 long position in CEI can be hedged for $0.0911 with a short position in Bitcoin. Moreover, in general, the hedging ratios are higher during COVID-19 than in the period before COVID-19 and continue to rise during the Russia-Ukraine war. This means that more cryptocurrencies are required when facing market shocks to reduce the risk of a green financial asset portfolio.

We also estimate the hedging effectiveness (HE) for the constructed portfolios during COVID-19 period and during Russia-Ukraine war period in Table 8. A higher value of HE signifies a more substantial reduction in portfolio risk, indicating an enhanced and more effective hedging strategy. Specifically, the hedging effectiveness of all seven cryptocurrencies in portfolios have increased during the Russia-Ukraine war period while the value in portfolios with XLM/ESG have decreased. This reflects the fact that cryptocurrencies can be used as a tool for portfolio diversification to hedge against risk for green financial markets when facing energy crises. Moreover, comparing the hedging effectiveness between conventional cryptocurrency and sustainable cryptocurrencies in portfolios with GBI, the estimation results of both the DCC-GARCH and Copula-GARCH models of sustainable cryptocurrencies are greater than those of BTC. This result holds during COVID-19 period and during Russia-Ukraine war periods. For example, the hedging effectiveness of XRP/GBI is 2.26% (DCC model) and 2.19% (Copula model) while the hedging effectiveness of BTC/GBI is 1.83% (DCC model) and 1.78% (Copula model), respectively. This interesting finding confirms that when facing market shocks, adding cryptocurrencies to a portfolio for diversification may be an effective option for investors to hedge against risks. In particular, some sustainable cryptocurrencies perform better in hedging than traditional cryptocurrencies, as they show higher hedging effectiveness in portfolios.

## 7. Discussion

Overall, the return connectedness spillover index indicates that when the COVID-19 and Russia-Ukraine war occurs, dynamic total connectedness reaches its peak. This outcome is consistent with prior studies showing that extreme economic events amplify intermarket linkages and spillovers (Diebold and Yilmaz 2012). Particularly, the short-term connectedness dominates both the medium-term and long-term connectedness for the whole period. The reason may be that prior to the shocks, markets were calmer, and investors were less fearful of market movements, in line with studies linking investor sentiment and uncertainty to spillover dynamics (Baker et al. 2016; Bouri et al. 2019). Shocks that create uncertainty in the system spread quickly and their impact on the system diminishes after a few days, thus leading to short-term connectedness. Considering the pairwise connectedness of return series, the lower degree of connectedness between cryptocurrencies and green financial market may provide diversification benefits in portfolio construction, where the same result can be found in Li and Meng (2022). Another interesting result is that the connectedness of sustainable cryptocurrencies to stocks is lower than the connectedness of Bitcoin to green financial market. As the off-diagonal results shown in Table 4, for instance, during the Russia–Ukraine war period, the pairwise connectedness between Bitcoin and the GBI was 0.28%, while that of XRP and the GBI was only 0.09%. This may reflect Bitcoin's larger market capitalization and broader institutional integration, leading to stronger spillovers, whereas sustainable cryptocurrencies remain less connected. Such weaker linkages suggest potential diversification benefits, consistent with prior evidence that assets with lower connectedness can enhance portfolio risk management (Mensi et al. 2021; Sharif et al. 2023). Therefore, this might encourage investors to pay extra attention to considering sustainable cryptocurrencies in their portfolios for hedging.

Regarding the volatility connectedness spillover index, the results are consistent with the connectedness of the return series. For instance, our dynamic analysis indicates that during the COVID-19 pandemic and the Russia–Ukraine war, overall volatility spillovers rise significantly, in agreement with study documenting that extreme events increase integration across volatility series (Zhang et al. 2020). Furthermore, following the commencement of the Russia-Ukraine war, the spillover effects of volatility between BTC and other sustainable cryptocurrencies have diminished, suggesting that traditional and sustainable cryptocurrencies do not bear the same risks when confronted with global crisis, as also noted in recent comparative studies (Duan et al. 2023).

The portfolio analysis results show that the two multivariate GARCH models exhibit some consistency in the estimation of hedging performance. The optimal hedging ratio tends to increase as market uncertainty grows. Specifically, the hedging ratios are higher during COVID-19 than in the period before COVID-19. It means that a higher proportion of cryptocurrencies are necessary in the portfolio during the COVID-19 period to mitigate the risk associated with green financial markets. Moreover, the hedging effectiveness of all cryptocurrencies in portfolios has increased significantly after the outbreak of Russia-Ukraine war. It reflects the fact that cryptocurrencies can be used as a tool for portfolio diversification to hedge against risk in the face of market shocks. When facing market shocks, adding cryptocurrencies to a portfolio for diversification may be an effective option for investors to hedge against risks. In particular, sustainable cryptocurrencies perform better in hedging the risk of green stock market than traditional cryptocurrencies, showing higher hedging effectiveness in portfolios, which corroborates recent findings on the distinctive risk–return profile of sustainable digital assets (Ali et al. 2024).

## 8. Conclusion

This study aims to examine the variation of information spillover as well as portfolio diversification between cryptocurrency and green financial markets over time. We employed the time-varying connectedness to represent the dynamic relationship of the return and volatility series in the period during the COVID-19 and Russia-Ukraine war. Our targets of observation contain traditional cryptocurrency, sustainable cryptocurrencies and green energy cryptocurrency. As for green financial markets, we adopt daily data of green bonds, green stocks and renewable energy indices. In order to investigate return and volatility connectedness between the two markets, we employ some multiple methods in this paper, such as the TVP-VAR model with Fourier transform, DCC-GARCH model and Copula-GARCH model. The results reveal that the correlations between cryptocurrencies and stocks related to green financial assets are relatively low, which is an essential finding for portfolio diversification. Afterwards, we choose distinct types of cryptocurrencies to build ideal portfolios including both traditional cryptocurrencies and sustainable cryptocurrencies.

In general, while there is no doubt that the process of mining cryptocurrencies has a negative impact on the environment, this may be mitigated if investors can choose to support the green financial market and sustainable cryptocurrencies while enjoying the profits that cryptocurrencies bring. Thus, investors can gain the diversification benefits of cryptocurrencies in the investment portfolios, and portfolio stability and ecological preservation do not necessarily contradict each other. By leveraging connectedness networks and hedging models, investors can monitor the dynamic correlations between cryptocurrencies and green financial assets, and promptly adjust their investment portfolios to optimize the risk-return ratio. Besides, policymakers can require cryptocurrency projects to disclose their energy consumption and environmental impact, and certify the projects that meet sustainable development standards, which help investors identify and support environmentally friendly cryptocurrencies. In future research, it is worth further exploring the risk management strategies of sustainable cryptocurrencies in portfolio. For instance, whether the risk diversification effects of sustainable cryptocurrencies could be increased using specific derivatives or innovative financial instruments in portfolio.

# Appendix

*1. Derivation of time-varying parameter vector autoregression (TVP-VAR)*

Specifically, the TVP-VAR model can be written in the following expressions:

$$\boldsymbol{Y}_t = \boldsymbol{\beta}_t \boldsymbol{Y}_{t-1} + \boldsymbol{\epsilon}_t \quad \boldsymbol{\epsilon}_t \mid \boldsymbol{F}_{t-1} \sim N(\boldsymbol{0}, \boldsymbol{S}_t) \tag{1}$$

$$\boldsymbol{\beta}_t = \boldsymbol{\beta}_{t-1} + \boldsymbol{\nu}_t \quad \nu_t \mid \boldsymbol{F}_{t-1} \sim N(\boldsymbol{0}, \boldsymbol{R}_t) \tag{2}$$

where $\boldsymbol{Y}_t$ represents a volatility vector of dimensions $N \times 1$, and $\boldsymbol{\beta}_t$ signifies a time-varying coefficient matrix with dimensions $N \times N_p$. The error vector $\boldsymbol{\epsilon}_t$ is $N \times 1$ in size, with a time-varying variance-covariance matrix denoted as $\boldsymbol{S}_t$ (with dimensions $N \times N$). Additionally, the parameter $\boldsymbol{\beta}_t$ is influenced by its own lagged value, $\boldsymbol{\beta}_{t-1}$. Moreover, $\boldsymbol{\nu}_t$ is an error disturbance vector of dimensions $N \times N_p$, accompanied by a time-varying variance-covariance matrix $\boldsymbol{R}_t$, which is $N_p \times N_p$ in size.

To facilitate this procedure, we convert the VAR into its vector moving average (VMA) representation:

$$\boldsymbol{Y}_t = \boldsymbol{\beta}_t \boldsymbol{Y}_{t-1} + \boldsymbol{\epsilon}_t \tag{3}$$

$$\boldsymbol{Y}_t = \boldsymbol{A}_t \boldsymbol{\epsilon}_t \tag{4}$$

$$\boldsymbol{A}_{0,t} = \boldsymbol{I} \tag{5}$$

$$\boldsymbol{A}_{i,t} = \boldsymbol{\beta}_{1,t} \boldsymbol{A}_{i-1,t} + \cdots + \boldsymbol{\beta}_{p,t} \boldsymbol{A}_{i-p,t} \tag{6}$$

where both $\boldsymbol{\beta}_{i,t}$ and $\boldsymbol{A}_{i,t}$ are parameter matrices with dimensions $N \times N$.

GIRF denotes the response of all variables after variable $i$ is shocked. Since the structural model is not provided, we calculate disparities between a J-step-ahead forecast in scenarios where variable $i$ is subject to a shock and where it is not. The dissimilarity in these forecasts can be attributed to the impact of the shock on variable $i$ and is computed as:

$$GIR_t(J, \boldsymbol{\delta}_{j,t}, \boldsymbol{F}_{t-1}) = E(\boldsymbol{Y}_{t+J} | \boldsymbol{\epsilon}_{j,t} = \boldsymbol{\delta}_{j,t}, \boldsymbol{F}_{t-1}) - E(\boldsymbol{Y}_{t+J} | \boldsymbol{F}_{t-1}) \tag{7}$$

$$\boldsymbol{\Psi}_{j,t}^{g}(J) = \frac{\boldsymbol{A}_{J,t}\boldsymbol{S}_t\boldsymbol{\epsilon}_{j,t}}{\sqrt{S_{jj,t}}} \tag{8}$$

where $J$ is the forecast horizon, and $\boldsymbol{F}_{t-1}$ denotes the information set until $t-1$. $\boldsymbol{\delta}_{j,t}$ is a selection vector, which equals to 1 at the jth position and 0 otherwise.

Subsequently, the Generalized Forecast Error Variance Decomposition (GFEVD) is determined by assessing the proportion of variance attributable to one variable in relation to the others. All variables collectively account for 100% of the forecast error variance for variable $i$. This computation is represented by the following equation:

$$\tilde{\phi}_{ij,t}^{g}(J) = \frac{\sum_{t=1}^{J-1} \Psi_{ij,t}^{2,g}}{\sum_{j=1}^{N}\sum_{t=1}^{J-1} \Psi_{ij,t}^{2,g}} \tag{9}$$

with $\sum_{j=1}^{N} \tilde{\phi}_{ij,t}^{N}(J) = 1$ and $\sum_{i,j=1}^{N} \tilde{\phi}_{ij,t}^{N}(J) = N$. By using the GFEVD, the *total connectedness index*, which shows how shocks to one variable spill over into other variables, can be derived as follows:

$$C_t^{g}(J) = \frac{\sum_{i,j=1,i\neq j}^{N} \tilde{\phi}_{ij,t}^{g}(J)}{N} \times 100. \tag{10}$$

Regarding the total directional connectedness to others, we initially examine the scenario where variable $i$ imparts its shock to all other variables, denoted as $j$:

$$C_{i\to j,t}^{g}(J) = \frac{\sum_{j=1,i\neq j}^{N} \tilde{\phi}_{ji,t}^{g}(J)}{\sum_{j=1}^{N} \tilde{\phi}_{ji,t}^{g}(J)} \times 100. \tag{11}$$

Furthermore, we compute the directed connectedness received by variable $i$ from variable $j$, called the total directional connectedness from others. This index is defined as follows:

$$C^{g}_{i\leftarrow j,t}(J) = \frac{\sum_{j=1,i\neq j}^{N} \tilde{\phi}^{g}_{ij,t}(J)}{\sum_{i=1}^{N} \tilde{\phi}^{g}_{ij,t}(J)} \times 100. \tag{12}$$

Finally, according to the Equation (11) and (12), the net total directional connectedness is derived. This difference can be interpreted as the impact of variable $i$ on the entire network of variables

$$C^{g}_{i,t} = C^{g}_{i\rightarrow j,t}(J) - C^{g}_{i\leftarrow j,t}(J). \tag{13}$$

A positive *net total directional connectedness* for variable $i$ suggests that it exerts more influence on the network than it receives from the network.

## 2. *Derivation of the connectedness network based on frequency decomposition*

The generalized variance decompositions on frequency band $d$, where $d = (a,b)\colon a,b \in (-\pi,\pi), a < b$, are defined as:

$$\theta_{i,j}(d) = \frac{1}{2\pi}\int_{d} \Gamma_i(\omega)(\mathbf{f}(\omega))_{i,j}\, \mathrm{d}\omega \tag{14}$$

where the weighting function $\Gamma_j(\omega)$ is

$$\Gamma_j(\omega) = \frac{(\Psi(e^{-i\omega})\Sigma\Psi'(e^{+i\omega}))_{j,j}}{\frac{1}{2\pi}\int_{-\pi}^{\pi}(\Psi(e^{-i\gamma})\Sigma\Psi'(e^{+i\gamma}))_{j,j}\, d\gamma} \tag{15}$$

and the generalized causation spectrum $\mathbf{f}(\omega)_{i,j}$ is

$$\mathbf{f}(\omega)_{i,j} = \frac{\sigma_{l,l}^{-1}|(\Psi(e^{-i\omega})\Sigma)_{i,j}|^2}{(\Psi(e^{-i\omega})\Sigma\Psi'(e^{+i\omega}))_{j,j}}. \tag{16}$$

Both $\omega \in (-\pi,\pi)$ and $\Psi(e^{-i\omega})$ are Fourier transforms of $\Psi_h$.

The within connectedness reflects the effect of connectedness within a frequency band and is derived by weighting the power of sequences on a particular frequency band. The equation of within connectedness is

$$\mathrm{S^M(d)} = 100 \times \frac{\sum_{i \neq j} \hat{\theta}_{ij}(d)}{\Sigma \hat{\theta}(d)} = 100 \times (1 - \frac{\mathrm{Tr}\{\hat{\theta}(d)\}}{\Sigma \hat{\theta}(d)}) \tag{17}$$

where $\mathrm{Tr}\{\cdot\}$ is the trace operator, and $\hat{\theta}(d)$ is the scaled generalized variance decomposition. The frequency connectedness is defined as:

$$\mathrm{S^F(d)} = (\frac{\sum \hat{\theta}(d)}{\sum \hat{\theta}(\infty)} - \frac{\mathrm{Tr}\{\hat{\theta}(d)\}}{\sum \hat{\theta}(\infty)}) \times 100 = \mathrm{S^M(d)} \times \frac{\Sigma \hat{\theta}(d)}{\Sigma \hat{\theta}(\infty)}. \tag{18}$$

## 3. *The total connectedness based on frequency decomposition*

Table 9 to Table 11 summarize the total connectedness of return series based on frequency decomposition (short-term, medium-term and long-term). The results show that the short-term connectedness are higher than medium- and long-term connectedness. And the pairwise connectedness between cryptocurrencies and green financial markets have remained at a low level across different frequency domains, although these values have increased in the face of market shocks.

Table 9: Short-term connectedness of return series

| | BTC | XRP | ADA | XLM | MIOTA | POWR | SNC | GBI | ESG | CEI | WIND | SOLAR | **FROM _ABS** | **FROM _WTH** |
|---|---|---|---|---|---|---|---|---|---|---|---|---|---|---|
| **Pre_COVID-19** | | | | | | | | | | | | | | |
| BTC | 22.86 | 10.19 | 11.66 | 9.85 | 10.13 | 9.10 | 5.88 | 0.01 | 0.00 | 0.00 | 0.01 | 0.02 | 4.74 | 6.09 |
| XRP | 10.10 | 20.87 | 13.14 | 11.89 | 11.46 | 9.14 | 3.79 | 0.01 | 0.22 | 0.22 | 0.07 | 0.16 | 5.02 | 6.45 |
| ADA | 10.19 | 11.66 | 18.93 | 12.42 | 11.49 | 10.48 | 4.29 | 0.00 | 0.07 | 0.07 | 0.06 | 0.12 | 5.07 | 6.52 |
| XLM | 9.56 | 11.58 | 13.78 | 20.66 | 10.33 | 10.33 | 4.74 | 0.04 | 0.10 | 0.10 | 0.07 | 0.13 | 5.06 | 6.51 |
| MIOTA | 9.54 | 11.19 | 12.62 | 10.26 | 21.17 | 10.09 | 5.81 | 0.01 | 0.04 | 0.04 | 0.03 | 0.03 | 4.97 | 6.39 |
| POWR | 8.98 | 9.28 | 12.00 | 10.45 | 10.40 | 21.98 | 6.89 | 0.06 | 0.15 | 0.16 | 0.10 | 0.11 | 4.88 | 6.27 |
| SNC | 8.29 | 5.15 | 7.21 | 7.45 | 9.09 | 10.95 | 34.70 | 0.02 | 0.10 | 0.11 | 0.22 | 0.07 | 4.06 | 5.21 |
| GBI | 0.06 | 0.30 | 0.37 | 0.09 | 0.07 | 0.19 | 0.06 | 70.15 | 0.59 | 0.61 | 4.94 | 0.65 | 0.66 | 0.85 |
| ESG | 0.28 | 0.39 | 0.31 | 0.48 | 0.09 | 0.36 | 0.16 | 0.33 | 27.12 | 26.98 | 8.91 | 8.16 | 3.87 | 4.97 |
| CEI | 0.27 | 0.38 | 0.30 | 0.46 | 0.09 | 0.35 | 0.16 | 0.32 | 27.02 | 27.10 | 9.12 | 8.22 | 3.89 | 5.00 |
| WIND | 0.46 | 0.76 | 0.70 | 0.61 | 0.26 | 0.47 | 0.38 | 2.16 | 11.46 | 11.73 | 32.87 | 6.63 | 2.97 | 3.81 |
| SOLAR | 0.37 | 0.59 | 0.39 | 0.41 | 0.06 | 0.26 | 0.11 | 0.30 | 11.80 | 11.90 | 7.71 | 39.22 | 2.83 | 3.63 |
| **TO_ABS** | 4.84 | 5.12 | 6.04 | 5.36 | 5.29 | 5.14 | 2.69 | 0.27 | 4.30 | 4.33 | 2.61 | 2.03 | 48.02 | |
| **TO_WTH** | 6.22 | 6.58 | 7.76 | 6.89 | 6.79 | 6.61 | 3.46 | 0.35 | 5.52 | 5.56 | 3.35 | 2.60 | | 61.70 |
| **During_COVID-19** | | | | | | | | | | | | | | |
| BTC | 19.58 | 6.50 | 10.33 | 8.76 | 10.23 | 7.12 | 6.66 | 0.94 | 3.72 | 3.77 | 2.93 | 2.72 | 5.31 | 6.58 |
| XRP | 8.09 | 23.89 | 10.06 | 14.07 | 10.88 | 5.91 | 3.74 | 0.38 | 2.02 | 2.04 | 1.10 | 1.05 | 4.94 | 6.13 |
| ADA | 10.90 | 8.48 | 20.21 | 11.81 | 11.86 | 6.50 | 4.45 | 0.62 | 3.06 | 3.08 | 1.95 | 1.78 | 5.37 | 6.66 |
| XLM | 9.50 | 12.02 | 11.54 | 20.15 | 10.88 | 6.45 | 4.62 | 0.52 | 2.33 | 2.36 | 1.71 | 1.33 | 5.27 | 6.54 |
| MIOTA | 10.33 | 8.66 | 11.19 | 10.50 | 20.08 | 8.53 | 4.81 | 0.37 | 2.82 | 2.83 | 1.91 | 1.79 | 5.31 | 6.59 |
| POWR | 9.17 | 5.78 | 7.47 | 7.45 | 10.87 | 27.02 | 4.93 | 0.22 | 2.61 | 2.63 | 1.86 | 1.52 | 4.54 | 5.63 |
| SNC | 10.93 | 5.59 | 7.02 | 7.47 | 7.81 | 6.18 | 32.88 | 0.41 | 2.25 | 2.29 | 1.58 | 1.16 | 4.39 | 5.44 |
| GBI | 1.86 | 0.54 | 1.34 | 0.89 | 0.76 | 0.38 | 0.51 | 35.89 | 6.36 | 6.49 | 7.74 | 3.90 | 2.56 | 3.18 |
| ESG | 4.03 | 1.63 | 3.16 | 2.33 | 3.04 | 2.25 | 1.47 | 2.64 | 21.42 | 21.33 | 10.94 | 8.12 | 5.08 | 6.30 |
| CEI | 4.05 | 1.65 | 3.17 | 2.36 | 3.06 | 2.27 | 1.49 | 2.69 | 21.21 | 21.23 | 11.11 | 8.27 | 5.11 | 6.34 |
| WIND | 3.30 | 0.94 | 2.09 | 1.83 | 2.02 | 1.62 | 1.12 | 4.39 | 11.62 | 11.87 | 22.31 | 11.67 | 4.37 | 5.42 |
| SOLAR | 3.66 | 1.16 | 2.25 | 1.73 | 2.18 | 1.69 | 1.03 | 1.92 | 10.48 | 10.72 | 13.47 | 25.46 | 4.19 | 5.20 |
| **TO_ABS** | 6.32 | 4.41 | 5.80 | 5.77 | 6.13 | 4.07 | 2.90 | 1.26 | 5.71 | 5.78 | 4.69 | 3.61 | 56.46 | |
| **TO_WTH** | 7.84 | 5.47 | 7.19 | 7.15 | 7.61 | 5.05 | 3.60 | 1.56 | 7.08 | 7.17 | 5.82 | 4.48 | | 70.02 |
| **During_R-U war** | | | | | | | | | | | | | | |
| BTC | 29.13 | 5.56 | 12.18 | 5.21 | 8.23 | 5.10 | 13.65 | 0.13 | 2.19 | 0.37 | 0.30 | 0.22 | 4.43 | 5.65 |
| XRP | 5.45 | 28.86 | 11.14 | 16.22 | 8.22 | 3.09 | 3.41 | 0.06 | 1.47 | 0.77 | 0.36 | 0.38 | 4.21 | 5.38 |
| ADA | 11.01 | 9.41 | 23.68 | 9.10 | 10.75 | 5.88 | 6.93 | 0.03 | 1.94 | 0.40 | 0.56 | 0.50 | 4.71 | 6.01 |
| XLM | 5.03 | 15.70 | 10.29 | 28.74 | 8.46 | 3.49 | 3.18 | 0.12 | 1.27 | 0.65 | 0.50 | 0.45 | 4.10 | 5.23 |
| MIOTA | 8.16 | 7.27 | 11.43 | 7.82 | 25.54 | 5.15 | 7.46 | 0.15 | 1.21 | 0.74 | 0.34 | 0.40 | 4.18 | 5.33 |
| POWR | 7.18 | 3.93 | 8.99 | 4.65 | 7.44 | 38.46 | 3.32 | 0.49 | 2.05 | 0.96 | 0.80 | 0.18 | 3.33 | 4.25 |
| SNC | 17.58 | 4.06 | 9.43 | 3.98 | 8.65 | 2.98 | 35.88 | 0.01 | 1.20 | 0.01 | 0.33 | 0.07 | 4.02 | 5.14 |
| GBI | 0.21 | 0.05 | 0.34 | 0.17 | 0.29 | 0.51 | 0.02 | 49.61 | 3.59 | 0.34 | 14.76 | 6.51 | 2.23 | 2.85 |
| ESG | 2.42 | 1.91 | 2.83 | 1.53 | 1.70 | 1.72 | 1.32 | 2.73 | 36.85 | 0.83 | 11.86 | 8.81 | 3.14 | 4.01 |
| CEI | 0.64 | 1.56 | 0.84 | 1.14 | 1.95 | 1.72 | 0.15 | 0.46 | 1.37 | 66.81 | 1.61 | 3.24 | 1.22 | 1.56 |
| WIND | 0.29 | 0.38 | 0.73 | 0.53 | 0.45 | 0.44 | 0.27 | 10.22 | 10.79 | 0.96 | 34.62 | 16.36 | 3.45 | 4.41 |
| SOLAR | 0.18 | 0.49 | 0.51 | 0.53 | 0.59 | 0.23 | 0.03 | 4.81 | 8.65 | 2.06 | 17.41 | 38.08 | 2.96 | 3.78 |
| **TO_ABS** | 4.85 | 4.19 | 5.73 | 4.24 | 4.73 | 2.53 | 3.31 | 1.60 | 2.98 | 0.67 | 4.07 | 3.09 | 41.99 | |
| **TO_WTH** | 6.19 | 5.35 | 7.31 | 5.41 | 6.04 | 3.22 | 4.23 | 2.04 | 3.80 | 0.86 | 5.19 | 3.95 | | 53.59 |

Notes: This table summarizes short-term connectedness of return series. The results depend on a 20-step-ahead GFEVD and a 1 lag TVP-VAR model (chosen by the minimum AIC value). The values on the diagonal represent the own return contribution of one asset and the values on the off diagonal represent the pairwise connectedness. TO_ABS and TO_WTH represent the total outflows

to absolute and the total outflows to within. FROM_ABS and FROM_WTH represent the total inflows from absolute and the total inflows from within.

Table 10: Medium-term connectedness of return series

| | BTC | XRP | ADA | XLM | MIOTA | POWR | SNC | GBI | ESG | CEI | WIND | SOLAR | **FROM _ABS** | **FROM _WTH** |
|---|---|---|---|---|---|---|---|---|---|---|---|---|---|---|
| **Pre_COVID-19** | | | | | | | | | | | | | | |
| BTC | 4.19 | 2.21 | 2.52 | 2.08 | 2.05 | 1.77 | 1.00 | 0.00 | 0.00 | 0.00 | 0.00 | 0.00 | 0.97 | 5.61 |
| XRP | 1.58 | 4.26 | 2.43 | 2.31 | 2.13 | 1.62 | 0.41 | 0.00 | 0.03 | 0.02 | 0.00 | 0.03 | 0.88 | 5.09 |
| ADA | 1.79 | 2.39 | 3.77 | 2.66 | 2.27 | 2.08 | 0.65 | 0.00 | 0.04 | 0.03 | 0.05 | 0.03 | 1.00 | 5.78 |
| XLM | 1.46 | 2.27 | 2.55 | 3.89 | 1.86 | 1.72 | 0.62 | 0.02 | 0.04 | 0.04 | 0.03 | 0.03 | 0.89 | 5.13 |
| MIOTA | 1.71 | 2.24 | 2.44 | 2.05 | 3.78 | 1.85 | 0.86 | 0.00 | 0.02 | 0.02 | 0.02 | 0.01 | 0.93 | 5.40 |
| POWR | 1.51 | 1.77 | 2.32 | 2.15 | 2.01 | 3.97 | 1.27 | 0.01 | 0.06 | 0.06 | 0.06 | 0.01 | 0.94 | 5.41 |
| SNC | 1.57 | 1.00 | 1.20 | 1.04 | 1.35 | 1.59 | 5.03 | 0.01 | 0.04 | 0.05 | 0.08 | 0.04 | 0.66 | 3.84 |
| GBI | 0.02 | 0.11 | 0.08 | 0.04 | 0.00 | 0.03 | 0.00 | 15.28 | 0.20 | 0.21 | 0.84 | 0.30 | 0.15 | 0.88 |
| ESG | 0.07 | 0.25 | 0.18 | 0.26 | 0.05 | 0.18 | 0.07 | 0.18 | 7.11 | 7.16 | 2.67 | 2.37 | 1.12 | 6.48 |
| CEI | 0.07 | 0.24 | 0.18 | 0.25 | 0.05 | 0.18 | 0.07 | 0.17 | 7.02 | 7.12 | 2.67 | 2.36 | 1.10 | 6.39 |
| WIND | 0.12 | 0.38 | 0.32 | 0.27 | 0.10 | 0.26 | 0.18 | 0.96 | 5.00 | 5.12 | 8.90 | 2.81 | 1.29 | 7.48 |
| SOLAR | 0.07 | 0.37 | 0.23 | 0.24 | 0.03 | 0.14 | 0.04 | 0.18 | 3.60 | 3.67 | 2.10 | 10.24 | 0.89 | 5.14 |
| **TO_ABS** | 0.83 | 1.10 | 1.20 | 1.11 | 0.99 | 0.95 | 0.43 | 0.13 | 1.34 | 1.37 | 0.71 | 0.67 | 10.83 | |
| **TO_WTH** | 4.80 | 6.37 | 6.96 | 6.44 | 5.73 | 5.49 | 2.50 | 0.74 | 7.73 | 7.90 | 4.11 | 3.86 | | 62.64 |
| **During_COVID-19** | | | | | | | | | | | | | | |
| BTC | 3.24 | 1.00 | 1.65 | 1.53 | 1.49 | 0.93 | 1.21 | 0.12 | 0.52 | 0.53 | 0.41 | 0.48 | 0.82 | 5.45 |
| XRP | 1.21 | 4.19 | 1.55 | 2.53 | 1.53 | 0.67 | 0.69 | 0.03 | 0.24 | 0.24 | 0.11 | 0.17 | 0.75 | 4.95 |
| ADA | 1.52 | 1.20 | 3.19 | 1.71 | 1.57 | 0.64 | 0.62 | 0.11 | 0.45 | 0.45 | 0.28 | 0.29 | 0.74 | 4.87 |
| XLM | 1.34 | 1.97 | 1.98 | 3.43 | 1.61 | 0.71 | 0.77 | 0.05 | 0.32 | 0.32 | 0.24 | 0.25 | 0.80 | 5.27 |
| MIOTA | 1.52 | 1.39 | 1.84 | 1.65 | 3.14 | 1.22 | 0.76 | 0.03 | 0.38 | 0.38 | 0.21 | 0.21 | 0.80 | 5.28 |
| POWR | 1.57 | 1.17 | 1.41 | 1.53 | 2.07 | 4.36 | 0.90 | 0.04 | 0.42 | 0.42 | 0.27 | 0.30 | 0.84 | 5.57 |
| SNC | 1.66 | 0.56 | 0.91 | 0.98 | 1.05 | 0.78 | 4.24 | 0.05 | 0.33 | 0.34 | 0.23 | 0.20 | 0.59 | 3.91 |
| GBI | 0.82 | 0.21 | 0.62 | 0.38 | 0.32 | 0.11 | 0.25 | 10.90 | 3.14 | 3.22 | 3.57 | 2.26 | 1.24 | 8.20 |
| ESG | 0.58 | 0.28 | 0.53 | 0.39 | 0.40 | 0.25 | 0.28 | 0.59 | 3.43 | 3.43 | 1.91 | 1.70 | 0.86 | 5.70 |
| CEI | 0.58 | 0.27 | 0.51 | 0.38 | 0.39 | 0.24 | 0.27 | 0.59 | 3.36 | 3.39 | 1.92 | 1.72 | 0.85 | 5.64 |
| WIND | 0.69 | 0.19 | 0.57 | 0.48 | 0.43 | 0.27 | 0.34 | 1.15 | 3.05 | 3.15 | 5.57 | 3.74 | 1.17 | 7.74 |
| SOLAR | 0.84 | 0.24 | 0.60 | 0.54 | 0.49 | 0.23 | 0.38 | 0.57 | 2.30 | 2.38 | 3.60 | 6.66 | 1.01 | 6.71 |
| **TO_ABS** | 1.03 | 0.71 | 1.01 | 1.01 | 0.94 | 0.50 | 0.54 | 0.28 | 1.21 | 1.24 | 1.06 | 0.94 | 10.48 | |
| **TO_WTH** | 6.79 | 4.67 | 6.70 | 6.67 | 6.25 | 3.34 | 3.56 | 1.84 | 8.00 | 8.19 | 7.04 | 6.24 | | 69.28 |
| **During_R-U war** | | | | | | | | | | | | | | |
| BTC | 4.50 | 0.90 | 2.34 | 0.78 | 1.78 | 1.16 | 2.17 | 0.02 | 0.19 | 0.02 | 0.02 | 0.00 | 0.78 | 4.62 |
| XRP | 1.18 | 5.61 | 2.50 | 2.92 | 1.82 | 0.58 | 0.71 | 0.03 | 0.36 | 0.07 | 0.16 | 0.13 | 0.87 | 5.16 |
| ADA | 1.74 | 1.95 | 5.11 | 1.68 | 2.30 | 1.13 | 1.10 | 0.01 | 0.33 | 0.02 | 0.09 | 0.02 | 0.86 | 5.11 |
| XLM | 1.07 | 3.30 | 3.23 | 5.60 | 2.04 | 0.84 | 0.65 | 0.06 | 0.19 | 0.04 | 0.18 | 0.07 | 0.97 | 5.75 |
| MIOTA | 1.49 | 2.45 | 3.50 | 1.88 | 6.56 | 1.31 | 1.02 | 0.05 | 0.29 | 0.14 | 0.10 | 0.13 | 1.03 | 6.10 |
| POWR | 1.49 | 0.97 | 2.73 | 1.16 | 1.89 | 7.31 | 0.88 | 0.03 | 0.17 | 0.18 | 0.02 | 0.01 | 0.79 | 4.70 |
| SNC | 2.12 | 0.72 | 1.63 | 0.57 | 1.65 | 0.52 | 4.97 | 0.00 | 0.19 | 0.00 | 0.02 | 0.00 | 0.62 | 3.66 |
| GBI | 0.05 | 0.03 | 0.14 | 0.05 | 0.12 | 0.07 | 0.00 | 11.35 | 1.40 | 0.07 | 3.74 | 1.36 | 0.59 | 3.47 |
| ESG | 0.60 | 0.59 | 0.87 | 0.49 | 0.61 | 0.32 | 0.26 | 0.79 | 9.76 | 0.37 | 3.05 | 2.10 | 0.84 | 4.96 |
| CEI | 0.16 | 0.23 | 0.09 | 0.20 | 0.26 | 0.24 | 0.01 | 0.21 | 0.29 | 11.50 | 0.47 | 0.83 | 0.25 | 1.48 |
| WIND | 0.10 | 0.10 | 0.28 | 0.13 | 0.18 | 0.12 | 0.06 | 2.87 | 3.09 | 0.16 | 7.95 | 3.61 | 0.89 | 5.27 |
| SOLAR | 0.03 | 0.13 | 0.16 | 0.17 | 0.17 | 0.01 | 0.00 | 1.82 | 3.23 | 0.37 | 5.33 | 9.09 | 0.95 | 5.64 |
| **TO_ABS** | 0.84 | 0.95 | 1.46 | 0.84 | 1.07 | 0.52 | 0.57 | 0.49 | 0.81 | 0.12 | 1.10 | 0.69 | 9.45 | |
| **TO_WTH** | 4.95 | 5.61 | 8.62 | 4.95 | 6.33 | 3.10 | 3.38 | 2.90 | 4.80 | 0.71 | 6.50 | 4.08 | | 55.93 |

Notes: This table summarizes medium-term connectedness of return series. The results depend on a 20-step-ahead GFEVD and a 1 lag TVP-VAR model (chosen by the minimum AIC value). The values on the diagonal represent the own return contribution of one asset and the values on the off diagonal represent the pairwise connectedness. TO_ABS and TO_WTH represent the total

outflows to absolute and the total outflows to within. FROM_ABS and FROM_WTH represent the total inflows from absolute and the total inflows from within.

Table 11: Long-term connectedness of return series

| | BTC | XRP | ADA | XLM | MIOTA | POWR | SNC | GBI | ESG | CEI | WIND | SOLAR | **FROM _ABS** | **FROM _WTH** |
|---|---|---|---|---|---|---|---|---|---|---|---|---|---|---|
| **Pre_COVID-19** | | | | | | | | | | | | | | |
| BTC | 1.16 | 0.62 | 0.71 | 0.59 | 0.57 | 0.50 | 0.28 | 0.00 | 0.00 | 0.00 | 0.00 | 0.00 | 0.27 | 5.58 |
| XRP | 0.43 | 1.19 | 0.67 | 0.64 | 0.59 | 0.45 | 0.11 | 0.00 | 0.01 | 0.01 | 0.00 | 0.01 | 0.24 | 4.96 |
| ADA | 0.50 | 0.67 | 1.06 | 0.75 | 0.64 | 0.58 | 0.18 | 0.00 | 0.01 | 0.01 | 0.01 | 0.01 | 0.28 | 5.75 |
| XLM | 0.40 | 0.64 | 0.71 | 1.09 | 0.52 | 0.48 | 0.17 | 0.01 | 0.01 | 0.01 | 0.01 | 0.01 | 0.25 | 5.05 |
| MIOTA | 0.48 | 0.63 | 0.68 | 0.58 | 1.05 | 0.52 | 0.24 | 0.00 | 0.01 | 0.01 | 0.01 | 0.00 | 0.26 | 5.35 |
| POWR | 0.42 | 0.50 | 0.65 | 0.60 | 0.56 | 1.10 | 0.35 | 0.00 | 0.02 | 0.02 | 0.02 | 0.00 | 0.26 | 5.36 |
| SNC | 0.44 | 0.28 | 0.34 | 0.29 | 0.37 | 0.44 | 1.38 | 0.00 | 0.01 | 0.01 | 0.02 | 0.01 | 0.19 | 3.80 |
| GBI | 0.01 | 0.03 | 0.02 | 0.01 | 0.00 | 0.01 | 0.00 | 4.29 | 0.06 | 0.06 | 0.23 | 0.09 | 0.04 | 0.86 |
| ESG | 0.02 | 0.07 | 0.05 | 0.08 | 0.01 | 0.05 | 0.02 | 0.05 | 2.03 | 2.05 | 0.77 | 0.68 | 0.32 | 6.59 |
| CEI | 0.02 | 0.07 | 0.05 | 0.07 | 0.01 | 0.05 | 0.02 | 0.05 | 2.01 | 2.04 | 0.76 | 0.68 | 0.32 | 6.49 |
| WIND | 0.03 | 0.11 | 0.09 | 0.08 | 0.03 | 0.08 | 0.05 | 0.29 | 1.46 | 1.50 | 2.55 | 0.82 | 0.38 | 7.75 |
| SOLAR | 0.02 | 0.11 | 0.07 | 0.07 | 0.01 | 0.04 | 0.01 | 0.05 | 1.03 | 1.05 | 0.59 | 2.91 | 0.25 | 5.20 |
| **TO_ABS** | 0.23 | 0.31 | 0.34 | 0.31 | 0.28 | 0.27 | 0.12 | 0.04 | 0.38 | 0.39 | 0.20 | 0.19 | 3.07 | |
| **TO_WTH** | 4.70 | 6.38 | 6.91 | 6.42 | 5.67 | 5.43 | 2.44 | 0.78 | 7.87 | 8.05 | 4.13 | 3.95 | | 62.73 |
| **During_COVID-19** | | | | | | | | | | | | | | |
| BTC | 0.90 | 0.28 | 0.46 | 0.43 | 0.41 | 0.25 | 0.34 | 0.03 | 0.14 | 0.14 | 0.11 | 0.14 | 0.23 | 5.35 |
| XRP | 0.33 | 1.16 | 0.43 | 0.70 | 0.42 | 0.18 | 0.19 | 0.01 | 0.06 | 0.06 | 0.03 | 0.05 | 0.20 | 4.82 |
| ADA | 0.41 | 0.33 | 0.88 | 0.47 | 0.43 | 0.17 | 0.17 | 0.03 | 0.12 | 0.13 | 0.08 | 0.08 | 0.20 | 4.74 |
| XLM | 0.37 | 0.54 | 0.55 | 0.95 | 0.44 | 0.19 | 0.21 | 0.01 | 0.09 | 0.09 | 0.07 | 0.07 | 0.22 | 5.16 |
| MIOTA | 0.41 | 0.38 | 0.51 | 0.45 | 0.86 | 0.33 | 0.21 | 0.01 | 0.10 | 0.10 | 0.06 | 0.05 | 0.22 | 5.13 |
| POWR | 0.43 | 0.33 | 0.39 | 0.43 | 0.58 | 1.20 | 0.25 | 0.01 | 0.12 | 0.12 | 0.08 | 0.09 | 0.23 | 5.52 |
| SNC | 0.46 | 0.15 | 0.25 | 0.27 | 0.29 | 0.21 | 1.16 | 0.01 | 0.09 | 0.09 | 0.06 | 0.06 | 0.16 | 3.82 |
| GBI | 0.24 | 0.06 | 0.18 | 0.11 | 0.09 | 0.03 | 0.08 | 3.16 | 0.92 | 0.95 | 1.06 | 0.68 | 0.37 | 8.62 |
| ESG | 0.16 | 0.08 | 0.15 | 0.11 | 0.11 | 0.07 | 0.08 | 0.17 | 0.96 | 0.96 | 0.54 | 0.49 | 0.24 | 5.71 |
| CEI | 0.16 | 0.07 | 0.14 | 0.11 | 0.11 | 0.06 | 0.08 | 0.17 | 0.94 | 0.95 | 0.54 | 0.49 | 0.24 | 5.63 |
| WIND | 0.19 | 0.05 | 0.16 | 0.14 | 0.12 | 0.07 | 0.10 | 0.33 | 0.87 | 0.90 | 1.59 | 1.08 | 0.33 | 7.88 |
| SOLAR | 0.24 | 0.06 | 0.17 | 0.15 | 0.14 | 0.06 | 0.11 | 0.17 | 0.66 | 0.68 | 1.04 | 1.92 | 0.29 | 6.84 |
| **TO_ABS** | 0.28 | 0.19 | 0.28 | 0.28 | 0.26 | 0.14 | 0.15 | 0.08 | 0.34 | 0.35 | 0.31 | 0.27 | 2.94 | |
| **TO_WTH** | 6.70 | 4.57 | 6.64 | 6.60 | 6.13 | 3.21 | 3.53 | 1.87 | 8.09 | 8.29 | 7.20 | 6.41 | | 69.22 |
| **During_R-U war** | | | | | | | | | | | | | | |
| BTC | 1.24 | 0.25 | 0.66 | 0.22 | 0.50 | 0.33 | 0.60 | 0.00 | 0.05 | 0.01 | 0.00 | 0.00 | 0.22 | 4.57 |
| XRP | 0.33 | 1.57 | 0.71 | 0.81 | 0.52 | 0.16 | 0.20 | 0.01 | 0.10 | 0.02 | 0.05 | 0.04 | 0.24 | 5.13 |
| ADA | 0.48 | 0.55 | 1.44 | 0.47 | 0.65 | 0.32 | 0.30 | 0.00 | 0.09 | 0.00 | 0.02 | 0.00 | 0.24 | 5.06 |
| XLM | 0.30 | 0.93 | 0.92 | 1.56 | 0.58 | 0.24 | 0.18 | 0.02 | 0.05 | 0.01 | 0.05 | 0.02 | 0.27 | 5.75 |
| MIOTA | 0.42 | 0.71 | 1.01 | 0.54 | 1.87 | 0.37 | 0.28 | 0.01 | 0.09 | 0.04 | 0.03 | 0.04 | 0.29 | 6.18 |
| POWR | 0.41 | 0.27 | 0.78 | 0.33 | 0.53 | 2.03 | 0.25 | 0.01 | 0.04 | 0.05 | 0.00 | 0.00 | 0.22 | 4.67 |
| SNC | 0.58 | 0.21 | 0.46 | 0.16 | 0.46 | 0.14 | 1.36 | 0.00 | 0.05 | 0.00 | 0.01 | 0.00 | 0.17 | 3.61 |
| GBI | 0.02 | 0.01 | 0.04 | 0.01 | 0.04 | 0.02 | 0.00 | 3.21 | 0.41 | 0.02 | 1.06 | 0.38 | 0.17 | 3.52 |
| ESG | 0.17 | 0.17 | 0.25 | 0.14 | 0.18 | 0.09 | 0.07 | 0.23 | 2.80 | 0.11 | 0.87 | 0.60 | 0.24 | 5.05 |
| CEI | 0.04 | 0.06 | 0.02 | 0.06 | 0.07 | 0.06 | 0.00 | 0.06 | 0.08 | 3.18 | 0.13 | 0.24 | 0.07 | 1.46 |
| WIND | 0.03 | 0.03 | 0.08 | 0.04 | 0.05 | 0.03 | 0.02 | 0.82 | 0.89 | 0.05 | 2.25 | 1.01 | 0.25 | 5.34 |
| SOLAR | 0.01 | 0.04 | 0.05 | 0.05 | 0.05 | 0.00 | 0.00 | 0.53 | 0.95 | 0.11 | 1.54 | 2.57 | 0.28 | 5.82 |
| **TO_ABS** | 0.23 | 0.27 | 0.41 | 0.23 | 0.30 | 0.15 | 0.16 | 0.14 | 0.23 | 0.03 | 0.31 | 0.19 | 2.68 | |
| **TO_WTH** | 4.88 | 5.67 | 8.70 | 4.93 | 6.34 | 3.08 | 3.31 | 2.97 | 4.92 | 0.71 | 6.60 | 4.06 | | 56.16 |

Notes: This table summarizes long-term connectedness of return series. The results depend on a 20-step-ahead GFEVD and a 1 lag TVP-VAR model (chosen by the minimum AIC value). The values on the diagonal represent the own return contribution of one asset and the values on the off diagonal represent the pairwise connectedness. TO_ABS and TO_WTH represent the total outflows

to absolute and the total outflows to within. FROM_ABS and FROM_WTH represent the total inflows from absolute and the total inflows from within.